\documentclass{iau} 
\usepackage{graphicx}
\usepackage{hyperref}

\title[The long-term evolution of stellar activity] 
{The long-term evolution of stellar activity}

\author[Scott G. Gregory]   
{Scott G. Gregory$^1$}

\affiliation{$^1$School of Physics \& Astronomy, University of St Andrews, St Andrews,
KY16 9SS, U. K. \\ email: {\tt sg64@andrews.ac.uk}}

\pubyear{2017}
\volume{328}  
\jname{Living around Active Stars}
\editors{D. Nandi, A. Valio \& P. Petit, eds.}

\begin{document}

\maketitle


\begin{abstract}
I review the evolution of low-mass stars with outer convective zones over timescales of millions-to-billions of years, from the pre-main sequence to solar-age, $\sim$4.6\,{\rm Gyr} (\cite[Bahcall et al. 1995]{bah95}; \cite[Amelin et al. 2010]{ame10}), and beyond.  I discuss the evolution of high-energy coronal and chromospheric emission, the links with stellar rotation and magnetism, and the emergence of the rotation-activity relation for stars within young clusters.
\end{abstract}


\firstsection 
\vspace{-3.5mm}
\section{Introduction}
The study of stellar activity encompasses observations at multiple wavelengths over a variety of timescales.  Analysis of stars of different age yields information about the long-term evolution of stellar activity, their magnetism, and about how the various layers of their atmospheres and their surface features change over time.  In a short review I can only scratch the surface of this vast field of several, complementary, research areas. I focus of a select few topics which are currently at the forefront of astrophysical research: the long-term evolution of stellar magnetism, and of coronal / chromospheric activity.  

In \S\ref{evolve}, I discuss the timescales for the main phases of stellar evolution, the pre-main sequence (PMS), the main sequence (MS), and the post-main sequence (post-MS), using a solar mass star as an example.  Stellar activity and rotation are intimately linked, as I discuss in \S\ref{rotation}. In \S\ref{PMSxray}, I discuss the evolution of the X-ray emission from PMS stars and the age at which we start to observe a rotation-activity relation. \S\ref{highenergy} deals with the long-term evolution, over Gyr, of the coronal and chromospheric activity, and the use of the latter as a proxy for stellar age.  Stellar activity diagnostics are related to the dynamo magnetic field generation process and the evolution of stellar magnetism. In \S\ref{magnetism}, I discuss the long-term evolution of stellar magnetic fields, and summarise in \S\ref{summary}.


\vspace{-3.5mm}
\section{The evolution of low-mass stars}\label{evolve}
The protostellar phase of embedded star formation is complete in $0.1-0.4\,{\rm Myr}$ (e.g. \cite[Dunham \& Vorobyov 2012]{dun12}; \cite[Offner \& McKee 2011]{off11}). Low-mass stars ($\sim0.1-2\,{\rm M}_\odot$) then enter the PMS phase as they contract under gravity, initially interacting magnetospherically with circumstellar disks.  A median disk lifetime is $\sim2-3\,{\rm Myr}$ (e.g. \cite[Haisch et al. 2001]{hai01}), which is a small fraction of the total PMS contraction time.


\subsection{Timescale for the PMS, MS, and post-MS phase}
Pre-main sequence stars are luminous because of the release of the gravitational potential energy as they contract.  Many PMS stars are more luminous than the Sun due to their large surface area ($L_\ast\propto R_\ast^2$). We can estimate the total PMS lifetime of a star by taking the ratio of available energy to the rate at which that energy is being used: in other words, the ratio of the gravitational energy to the stellar luminosity,
\begin{equation}
\tau_{\rm PMS} = \frac{G M_\ast^2}{R_\ast L_\ast} \approx 3\times10^7 \left(\frac{M_\ast}{{\rm M}_\odot}\right)^2\left(\frac{R_\ast}{{\rm R}_\odot}\right)^{-1}\left(\frac{L_\ast}{{\rm L}_\odot}\right)^{-1} \hspace{2mm} {\rm years},
\label{tau_pms}
\end{equation}
where the symbols have obvious meaning.  $\tau_{\rm PMS}$ is the Kelvin-Helmholtz timescale. A solar mass stars takes $\sim30\,{\rm Myr}$ to complete its PMS contraction and reach the ZAMS.  

Once a star has reached the MS we can estimate how long it will take to exhaust its supply of hydrogen fuel - this is the nuclear timescale.  Similar to the timescale for PMS contraction, the MS lifetime is obtained by taking the ratio of energy available to the luminosity, although in this case reduced by the fraction of the stellar interior over which the nuclear reactions occur.  In the fusion of H into He about $0.7\%$ of the mass is released as energy (heat), with reactions occurring within the inner $10\%$ only for a solar mass star (\cite[Maeder 2009]{mae09}). Therefore, the amount of energy available from hydrogen fusion is roughly $7\times10^{-4}M_\ast c^2$, giving an approximate main sequence lifetime of, 
\begin{equation}
\tau_{\rm MS} \approx 7\times10^{-4}\frac{M_\ast c^2}{L_\ast} \approx 10^{10}\left(\frac{M_\ast}{{\rm M}_\odot}\right)\left(\frac{L_\ast}{{\rm L}_\odot}\right)^{-1} \hspace{2mm} {\rm years}.
\label{tau_ms}
\end{equation}
A solar mass star spends approximately $10\,{\rm Gyr}$ on the main sequence and $\tau_{\rm MS}\gg\tau_{\rm PMS}$.

Post-main sequence evolution is highly dependent on the stellar mass and metallicity. A solar mass star evolves through the subgiant phase as it turns off the MS and onto the red giant branch in the Hertzsprung-Russell (H-R) diagram. H burning continues in a shell surrounding the extinct He core as the star expands.  The He flash occurs as the star reaches the tip of the red giant branch and He fusion begins. A solar mass star then moves onto the horizontal branch (or to a region of the H-R diagram known as the red clump if of solar metallicity). Next is the asymptotic giant branch phase as He continues burning in a shell around the core, and eventually the planetary nebula phase which leaves behind a white dwarf. The total timescale for the post-MS evolution of a $1\,{\rm M}_\odot$ and solar metallicity star is roughly $\tau_{\rm postMS}\approx3.3\,{\rm Gyr}$ (\cite[Pols et al. 1998]{pol98}) with $\tau_{\rm postMS}<\tau_{\rm MS}$ and $\tau_{\rm postMS}\gg\tau_{\rm PMS}$.        


\subsection{The evolution of low-mass stars in human terms}
As astronomers we are used to working with stellar ages of order millions, and billions, of years.  However, it is close to impossible for us to fully comprehend the immensity of such numbers, unless we rescale them to something commensurate with our life experiences.     

In the United Kingdom the life expectancy of a newborn girl (boy) is 82.8 (79.1) years, assuming mortality rates remain the same as they were between 2013 and 2015 throughout their lives.\footnote{Office for National Statistics licensed under the Open Government Licence v.3.0.}  Let's assume that a person will live to be 80 years of age.  From formation to ending up as a white dwarf takes $\sim$13.33$\,{\rm Gyr}$ for a solar mass star, see \S\ref{evolve}. Of this total lifetime $\sim$30$\,{\rm Myr}$ ($\sim$0.2\%) is the duration of the PMS phase, $\sim$10$\,{\rm Gyr}$ ($\sim$75\%) the duration of the MS phase, and $\sim$3.3$\,{\rm Gyr}$ ($\sim$24.8\%) the duration of the post-MS evolution. This means, that for an 80 year life expectancy, the PMS phase is complete in only $\sim$9 weeks. At 9 weeks of age babies cannot bring their hands to their mouths (\cite[Gerber et al. 2010]{ger10}).  The MS phase, in human terms, is complete by an age of $\sim$60 years.  This means that the remaining $\sim$20 years of the life of our average person, mainly the retirement years, correspond to the post-MS evolution of a solar mass star.             

\begin{figure}[t]
\begin{center}
 \includegraphics[width=0.53\linewidth]{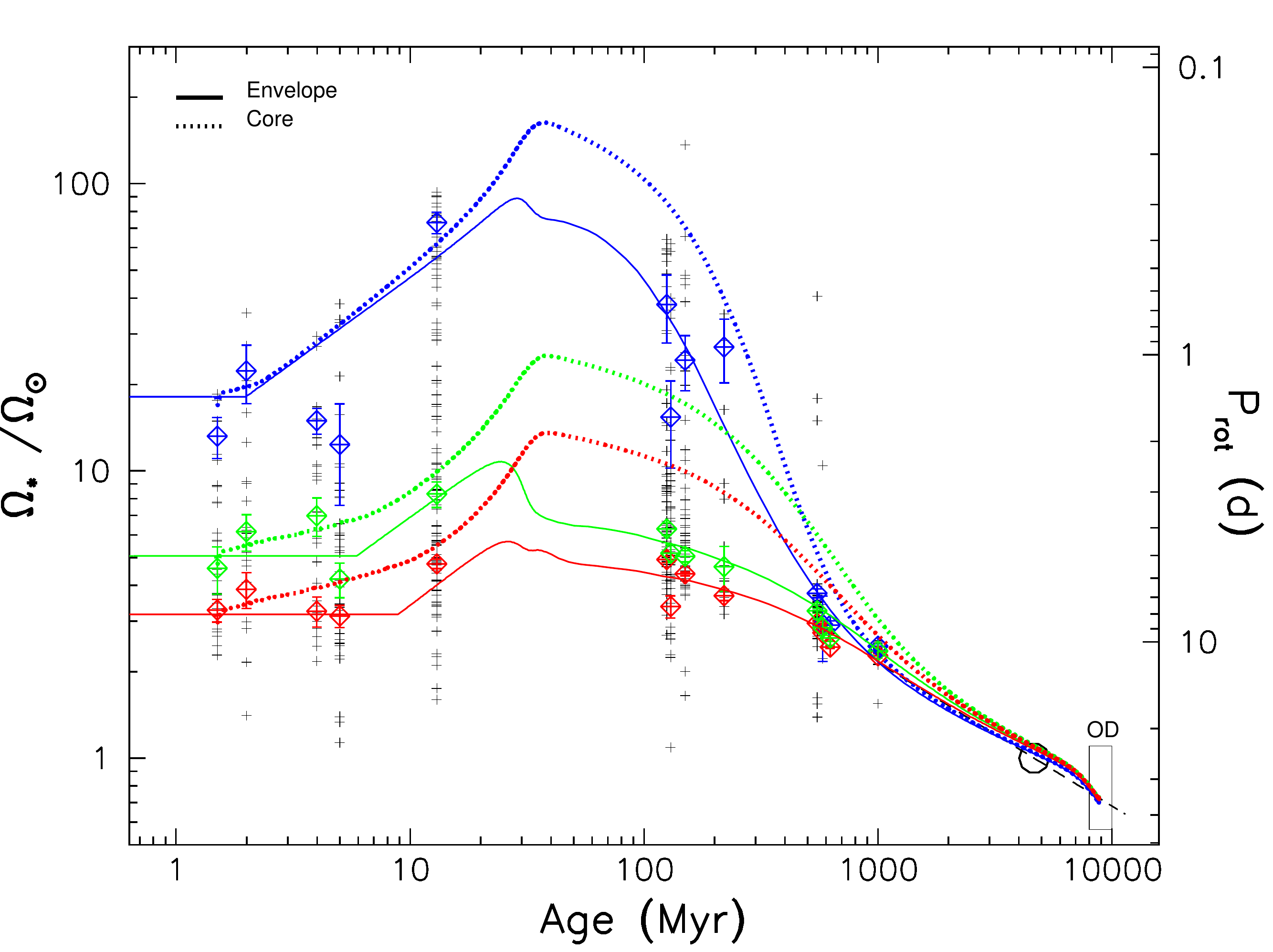} 
 \caption{The rotation rate of solar mass stars normalised to the solar rotation rate, $\Omega_\ast/\Omega_\odot$, versus age, from \cite[Gallet \& Bouvier (2015)]{gal15}. Points are rotation rates of stars in clusters of different age. The red/green/blue points represent the lower quartile/median/90th percentile of the rotation rate distribution in each cluster. The solid/dotted lines are models showing the rotation rate variation of the convective envelope/radiative core.  $\Omega_\ast$ is assumed constant during the disk-locked phase, with stars then spinning up as they complete their PMS contraction.  Stars then spin down, and rotation rates converge, through angular momentum lose in magnetised winds.  Credit: Gallet \& Bouvier, A\&A, 577, A98, 2015, reproduced with permission \copyright ESO.}
   \label{omega_age}
\end{center}
\end{figure}

When framed in human terms, the brevity of the PMS phase is readily apparent.  However, the PMS evolution sets the initial conditions for the subsequent evolution of the star.  As one example, Figure \ref{omega_age}, from \cite[Gallet \& Bouvier (2015)]{gal15}, illustrates the theorised behaviour of the stellar rotation rate $\Omega_\ast$ as a function of stellar age.  Following a period of disk-locking, where it is assumed that PMS stars are losing enough angular momentum that they accrete from their disks and contract while maintaining a constant rotation period, they are free to spin-up as they conserve angular momentum for their remaining PMS evolution (e.g. \cite[Davies et al. 2014]{dav14}).  Within young clusters there remains a considerable range of rotation rates until an age of $\sim0.5\,{\rm Gyr}$. Stars spin down via angular momentum lose from magnetised winds, and as the spin down rate is greatest for the faster rotating stars, there is convergence in the spread of rotation rates.  Beyond approximately $0.5\,{\rm Gyr}$ stars are spinning down with age as $\Omega_\ast\propto t^{-1/2}$, the Skumanich spin-down law \cite[(Skumanich 1972)]{sku72}. This convergence in rotation rates corresponds to a reduction in the scatter of activity levels with age, as I discuss in the following sections.


\vspace{-5mm}
\section{Stellar activity and rotation}\label{rotation}
It has long been known that rotation and activity are linked. \cite[Skumanich (1972)]{sku72} demonstrated that the decay in chromospheric CaII H \& K emission with increasing age was coupled with a decrease in stellar rotation rate, $\Omega_\ast\propto t^{-1/2}$.   \cite[Pallavicini et al. (1981)]{pal81} reported that the coronal activity of main sequence stars scales with rotation, with X-ray luminosities $L_{\rm X}\propto (v_\ast \sin i)^2$.  If we take from dynamo theory that dynamos are linear with the magnetic field $B\propto \Omega_\ast$ [roughly what is found from magnetic mapping studies, see \cite[Vidotto et al. (2014)]{vid14} and \S\ref{magnetism}], then this implies that $L_{\rm X}\propto B^2$; that is, that the X-ray luminosity scales with the magnetic energy density.  We may have expected this given that X-ray emission is the ultimate consequence of the magnetic activity and reconnection.         

\begin{figure}[t]
\begin{center}
 \includegraphics[width=0.5\linewidth]{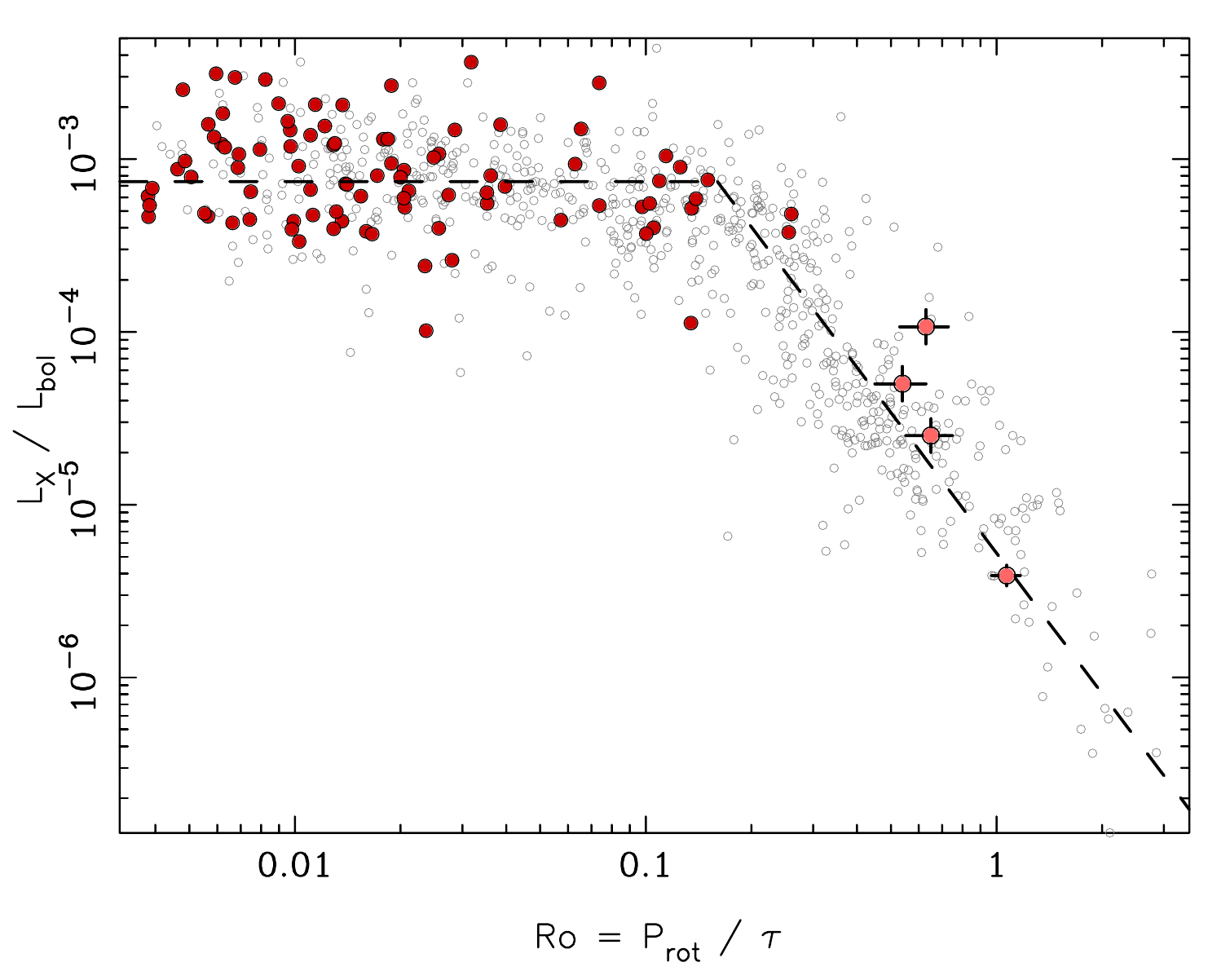} 
 \caption{The rotation-activity relation for MS stars - the fractional X-ray luminosity versus Rossby number, from \cite[Wright \& Drake (2016)]{wri16}. Grey points are stars with partially convective interiors. Red points are fully convective stars, four of which (with error bars on their positions) lie within the unsaturated regime. Reprinted by permission from Macmillan Publishers Ltd: Nature (Wright \& Drake 2016, Nature, 535, 526-528), copyright 2016.}
   \label{wright}
\end{center}
\end{figure}

X-ray emission from MS stars does not increase ad infinitum with increasing rotation rate.  For rapidly rotating stars the fractional X-ray luminosity, $L_{\rm X}/L_\ast$, saturates at about $10^{-3}$. Figure \ref{wright}, from \cite[Wright et al. (2016)]{wri16}, shows the X-ray rotation-activity relation for MS stars, with the unsaturated (where $L_{\rm X}/L_\ast$ decreases with decreasing rotation rate) and saturated (where $L_{\rm X}/L_\ast$ is roughly constant) regimes highlighted.\footnote{A third regime, super-saturation, exists where $L_{\rm X}/L_\ast$ decays for the most rapid rotators, likely as a result of coronal stripping (\cite[Jardine \& Unruh 1999]{jar99}; \cite[James et al. 2000]{jam00}; \cite[Jardine 2004]{jar04}).}  Rotation-activity relations, first quantified by \cite[Noyes et al. (1984)]{noy84} using chromospheric CaII H \& K emission, are also found for other activity diagnostics, such as H$\alpha$ emission (e.g. \cite[Soderblom et al. 1993]{sod93}; \cite[Douglas et al. 2014]{dou14}; \cite[Newton et al. 2016]{new16}), flare activity (\cite[Audard et al. 2000]{aud00}; \cite[Davenport 2016]{dav16}), the magnetic flux $Bf$, where $f$ is the magnetic filling factor (\cite[Reiners et al. 2009]{rei09}), and surface averaged magnetic field strengths (\cite[Vidotto et al. 2014]{vid14}).  

Rotation-activity relations are usually plotted with the Rossby number, $Ro$, as abscissae values, where $Ro=P_{\rm rot}/\tau_{\rm c}$ is the ratio of rotation period to the convective turnover time of gas cells within the stellar interior.  Plotting against Rossby number does tighten rotation-activity relations compared to plotting against rotation period alone.  However, this is often by design, with (model dependent) convective turnover times calibrated to minimise scatter for stars that fall in the unsaturated regime (e.g. \cite[Pizzolato et al. 2003]{piz03}).  It is not certain if ${\rm Ro}$ is the best quantity to use for rotation-activity studies, or if the rotation period itself is sufficient (\cite[Reiners et al. 2014]{rei14}).      

The cause of the saturation of activity diagnostics is still debated.  Multiple ideas have been proffered. It may be due to the stellar surface becoming completely filled with active regions (\cite[Vilhu 1984]{vil84}).  This is unlikely, however, as rotationally modulated X-ray emission has been detected from saturated regime stars [from young PMS stars at least, which all lie in the saturated regime e.g. \cite[Flaccomio et al. (2005)]{fla05} and \S\ref{PMSxray}], suggesting that there are X-ray dark regions, analogous to solar coronal holes, the location of stellar wind-bearing open field lines.  Other causes may be the saturation of the underlying dynamo itself, or centrifugal stripping of the corona (\cite[Jardine \& Unruh 1999]{jar99}). 

A star's position in the rotation-activity plane depends on multiple parameters. Age, rotation rate, mass, spectral type, stellar internal structure, and magnetic field topology are all interlinked factors.  In particular, the rotation-activity relation plotted in Figure \ref{wright} contains a mixture of fully convective stars (red points) and partially convective stars with radiative zones and outer convective envelopes (grey points).  Saturation occurs at $L_{\rm X}/L_\ast\approx 10^{-3}$, regardless of spectral type (e.g. \cite[Vilhu 1984]{vil84}), and for ${\rm Ro}\lesssim0.1$.  Therefore, the rotation period for which saturation occurs is longer for later spectral type stars as they have larger convective turnover times (e.g. \cite[Landin et al. 2010]{lan10}; \cite[Wright et al. 2011]{wri11}).  

MS stars of spectral type $\sim$M3.5 and later have fully convective interiors and are expected and found to show saturated levels of activity.  Slowly rotating earlier spectral type stars have partially convective interiors and follow the unsaturated regime with their activity level depending on their rotation rate. However, \cite[Wright et al. (2016)]{wri16} have recently demonstrated that fully convective M-dwarfs, at least those which have sufficiently spun down, also follow the unsaturated regime, see Figure \ref{wright}.  This behaviour was hinted at previously in data published by \cite[Kiraga \& Stepien (2007)]{kir07}, and \cite[Jeffries et al. (2011)]{jef11} who split the rotation activity-relation by stellar mass.  In their lowest mass sample, stars of mass $<0.35\,{\rm M}_\odot$ (which is the mass below which MS stars have fully convective interiors; \cite[Chabrier \& Baraffe 1997]{cha97}), it can be seen that 4 stars display unsaturated rotation-activity behaviour.  However, the stars considered by \cite[Jeffries et al. (2011)]{jef11} lie close to the fully convective boundary, and 3 may be partially convective or saturated.  The fully convective stars studied by \cite[Wright et al. (2016)]{wri16} that fall in the unsaturated regime are of sufficiently late spectral type that they are almost certainly fully convective.  Therefore, the existence of a tachocline between a radiative core and outer convective envelope is not required in order to generate stellar magnetic fields that link the activity level with rotation rate.       


\vspace{-3.5mm}
\section{The emergence of the rotation-activity relation on the PMS}\label{PMSxray}
\begin{figure}[t]
\begin{center}
 \includegraphics[width=0.337\linewidth]{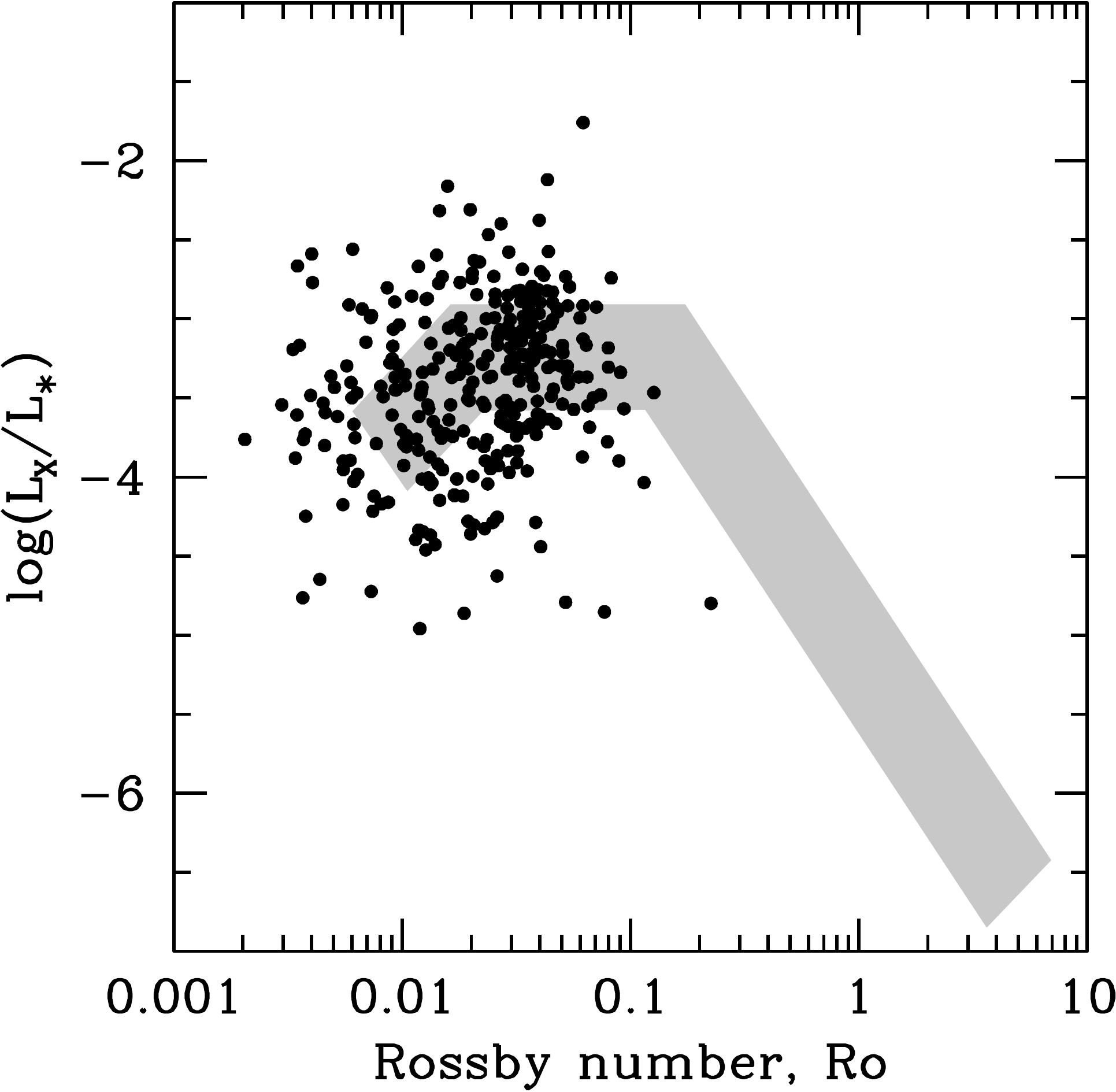} 
 \includegraphics[width=0.50\linewidth]{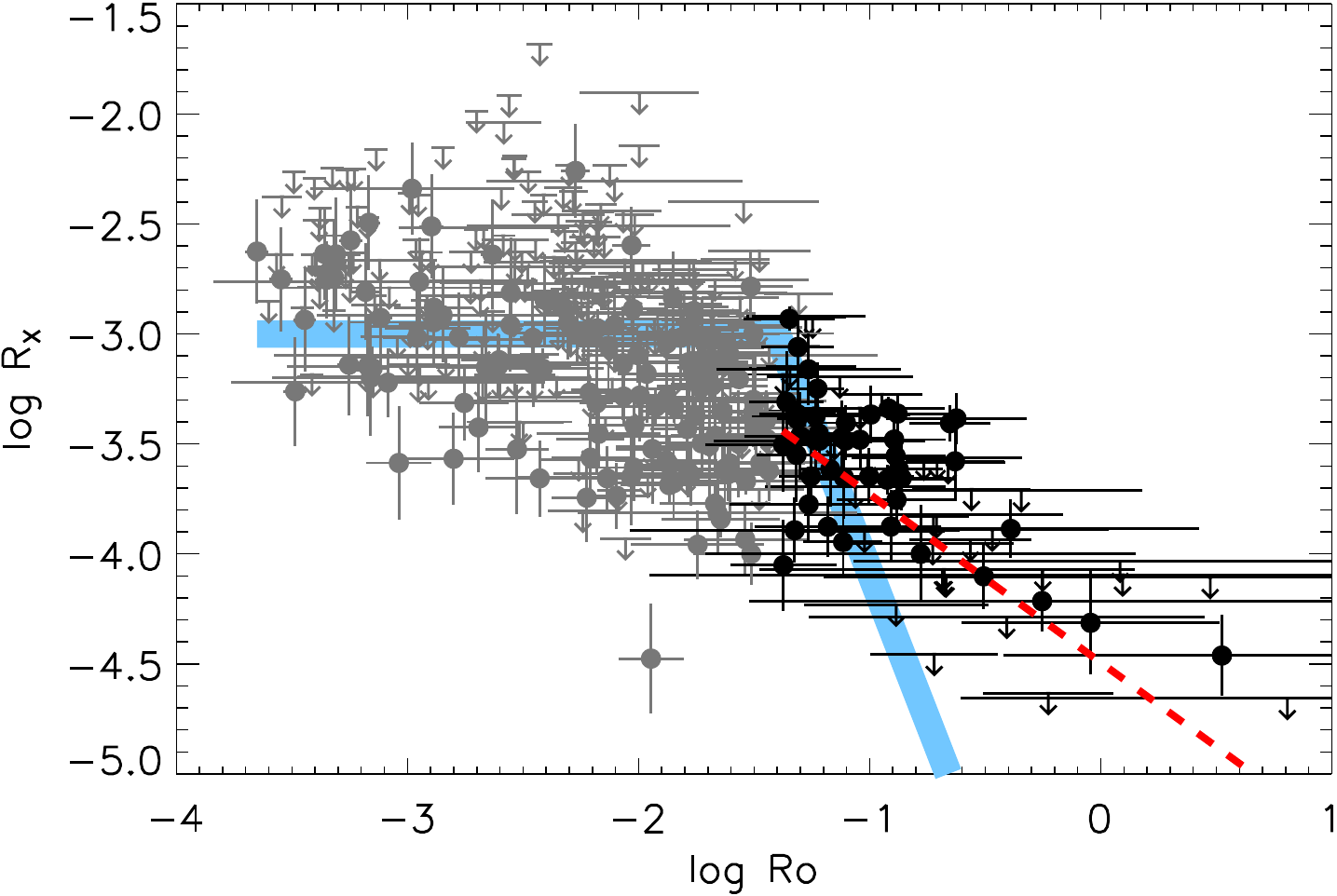} 
 \caption{X-ray rotation-activity plots for PMS stars. (left) Stars in the young Orion Nebula Cluster (ONC) and IC~348 star forming regions (black points).  The MS rotation-activity relation, with (left-to-right) the super-saturated, saturated, and unsaturated regimes is represented by the grey box. PMS stars show saturated levels of X-ray emission with far more scatter in $L_{\rm X}/L_\ast$ compared to saturated MS stars.  Figure from Gregory et al. (in prep.) using X-ray data from \cite[Stelzer et al. (2012)]{ste12} and \cite[Broos et al. (2013)]{bro13}. (right) PMS stars (points) in the $\sim13\,{\rm Myr}$ old cluster h~Per ($R_{\rm X}\equiv L_{\rm X}/L_\ast$) from \cite[Argiroffi et al. (2016)]{arg16}. The solid blue line represents the saturated and unsaturated regime for MS stars, from \cite[Wright et al. (2011)]{wri11}. The dashed red line is the fit to h~Per members expected to be in the unsaturated regime (black points).  Credit: Argiroffi et al., A\&A, 589, A113, 2016, reproduced with permission \copyright ESO.}
   \label{my_rotactivity}
\end{center}
\end{figure}

PMS stars in the youngest star forming regions do not follow the rotation-activity relation. PMS stars show saturated levels of X-ray emission regardless of their rotation rates, but with orders of magnitude more scatter in $L_{\rm X}/L_\ast$ compared to saturated regime MS stars (\cite[Preibisch et al. 2005]{pre05a}; \cite[Alexander \& Preibisch 2012]{ale12}; Figure \ref{my_rotactivity} left panel).  The origin of this scatter is intrinsic differences between PMS stars as it is greater than what can be attributed to observational uncertainties in the parameters (\cite[Preibisch et al. 2005]{pre05a}).   

The scatter in the rotation-activity plane for stars in young PMS clusters must eventually evolve to form the MS rotation-activity relation.  Recently, \cite[Argiroffi et al. (2016)]{arg16} have reported that PMS stars (at least those of mass $>1\,{\rm M}_\odot$) in the $\sim13\,{\rm Myr}$ cluster h~Per have begun to display activity regimes like MS stars. This can be seen in Figure \ref{my_rotactivity} (right panel). For h~Per the gradient of the fit to the unsaturated regime PMS stars is noticeably less when compared to the equivalent MS stars.  By $\sim13\,{\rm Myr}$ PMS coronal activity has evolved from the scatter evident for younger clusters (see Figure \ref{my_rotactivity}, left panel) to beginning to show a link with rotation, at least for stars more massive than a solar mass.  This may be due to the establishment of ``solar-type'' $\alpha\Omega$-dynamos, with the development of a shear layer, the tachocline, between an inner radiative core and an outer convective envelope within stellar interiors. 

The age of h~Per, $\sim13\,{\rm Myr}$, is (in rough terms) the age at which a solar-mass star makes the transition to the Henyey track in the Hertzsprung-Russell diagram (e.g. \cite[Siess et al. 2000]{sie00}).  By the time it does so, the stellar interior is already mostly radiative with the core containing the majority of the stellar mass (\cite[Gregory et al. 2016]{gre16}). The stellar internal structure transition, from fully to partially convective, and particularly when PMS stars evolve from the Hayashi track to the Henyey track, corresponds to an increase in the complexity of their large-scale magnetic field topology (\cite[Gregory et al. 2012, 2016]{gre12,gre16}). Hayashi track PMS stars (at least those of mass $\gtrsim0.5\,{\rm M}_\odot$) have simple axisymmetric large-scale magnetic fields which evolve into complex, multipolar, and dominantly non-axisymmetric magnetic fields once they develop large radiative cores and evolve onto Henyey tracks.  This evolution from the Hayashi to Henyey track also corresponds to a decay of the coronal X-ray emission (\cite[Rebull et al. 2006]{reb06}; \cite[Gregory et al. 2016]{gre16}).  
    
\begin{figure}[t]
\begin{center}
 \includegraphics[width=0.32\linewidth]{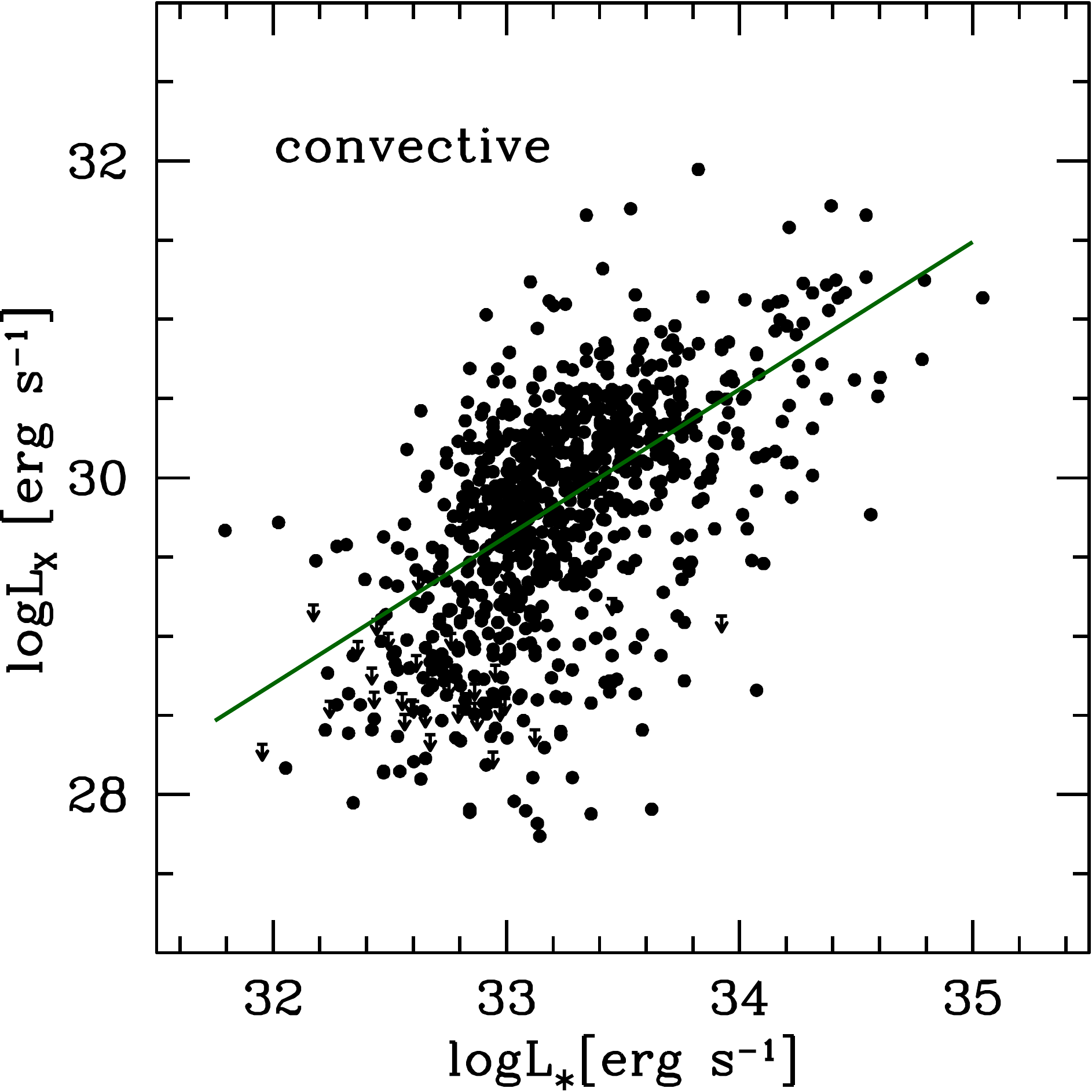} 
 \includegraphics[width=0.32\linewidth]{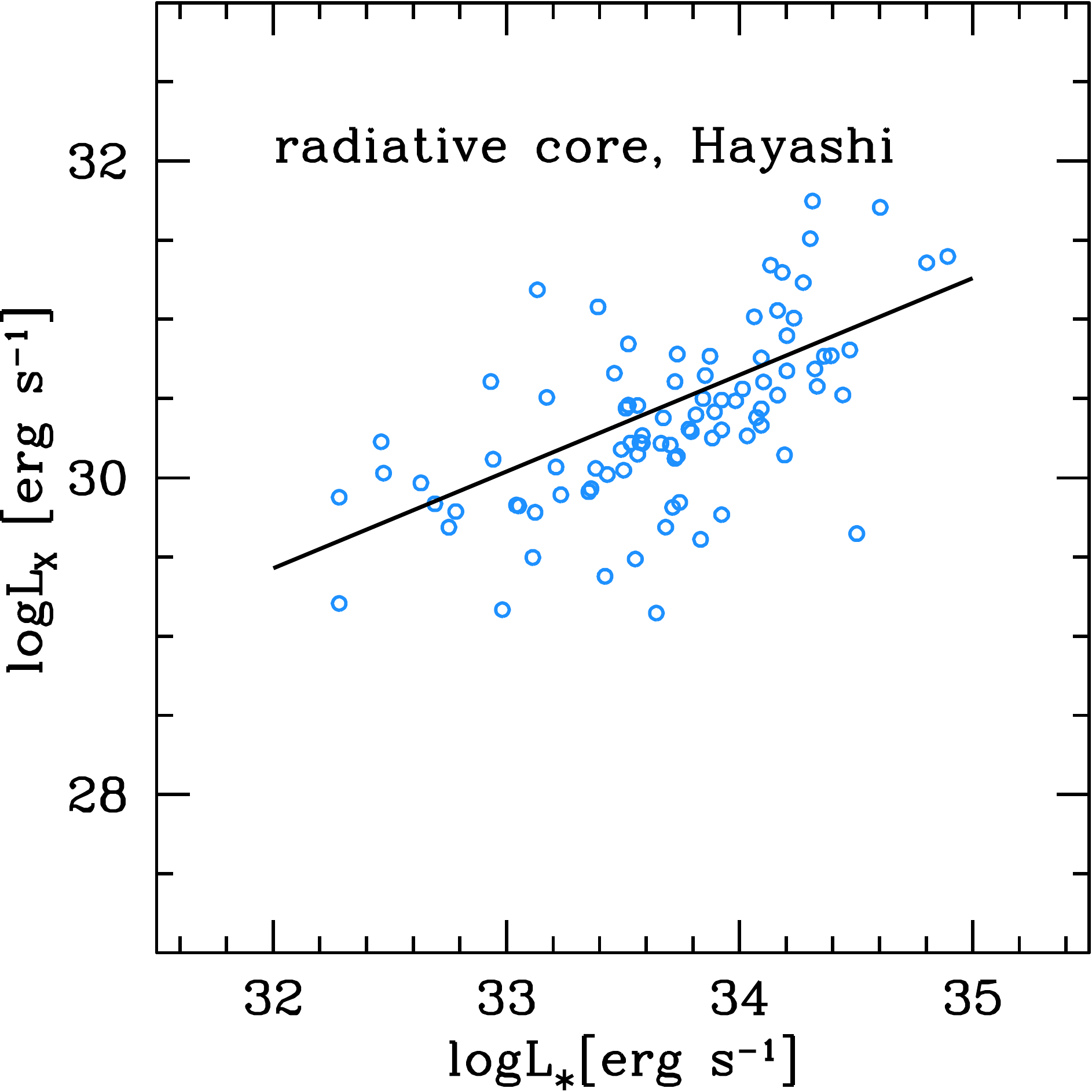}
 \includegraphics[width=0.32\linewidth]{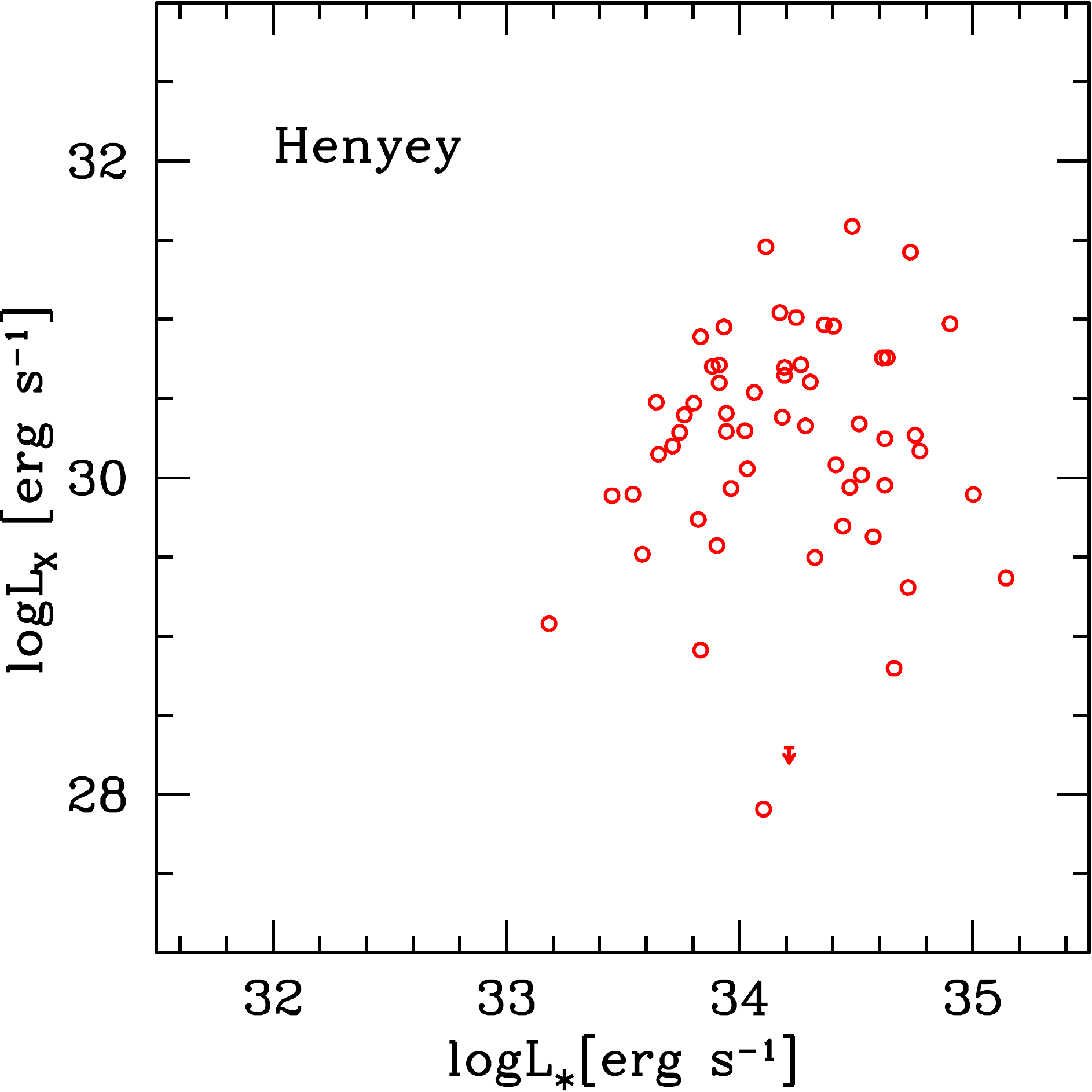}
 \caption{$L_{\rm X}$ and $L_\ast$ are correlated for Hayashi track PMS stars (left / middle) but there is no correlation for Henyey track PMS stars (right), from Gregory et al. (2016). With $L_{\rm X}\propto L_\ast^a$, $a=0.93\pm0.04$ for fully convective PMS stars with the exponent reducing to $a=0.61\pm0.08$ for partially convective Hayashi track PMS stars (those with small radiative cores).}
   \label{logLX_logLstar}
\end{center}
\end{figure}    
    
Pre-main sequence stars that have evolved onto Henyey tracks have lower $L_{\rm X}/L_\ast$ compared to those on Hayashi tracks (\cite[Rebull et al. 2006]{reb06}; \cite[Gregory et al. 2016]{gre16}).  Henyey track stars are also, on average, more luminous than Hayashi track stars.  Therefore, is the reduction in $L_{\rm X}/L_{\ast}$ driven mainly by the increase in $L_\ast$ or a decay of $L_{\rm X}$ when PMS stars evolve onto Henyey tracks?  It can be seen from Figure \ref{logLX_logLstar} that the coronal X-ray emission decays with radiative core development, and especially once PMS stars have evolved onto Henyey tracks. Considering a sample of almost 1000 PMS stars in 5 young star forming regions (\cite[Gregory et al. 2016]{gre16}), there is an almost linear relationship with $L_{\rm X}\propto L_\ast^{0.93\pm0.04}$ for fully convective PMS stars.  This is not surprising given that they all fall in the saturated regime of the rotation-activity plane where $L_{\rm X}/L_\ast \approx\,{\rm const.}$ (see Figure \ref{my_rotactivity}). For stars that have begun to develop radiative cores, but which are still on Hayashi tracks (where $L_\ast$ is decreasing with increasing age) the exponent is less with $L_{\rm X}\propto L_\ast^{0.61\pm0.08}$.  In both cases, the probability of there not being a correlation is $<5{\rm e}$-$5$ from a generalised Kendall's $\tau$ test. For Henyey track PMS stars, those with substantial radiative cores, there is no correlation between $L_{\rm X}$ and $L_\ast$, with many having a X-ray luminosity well below what would be expected for their bolometric luminosity.

\begin{figure}[t]
\begin{center}
 \includegraphics[width=0.31\linewidth]{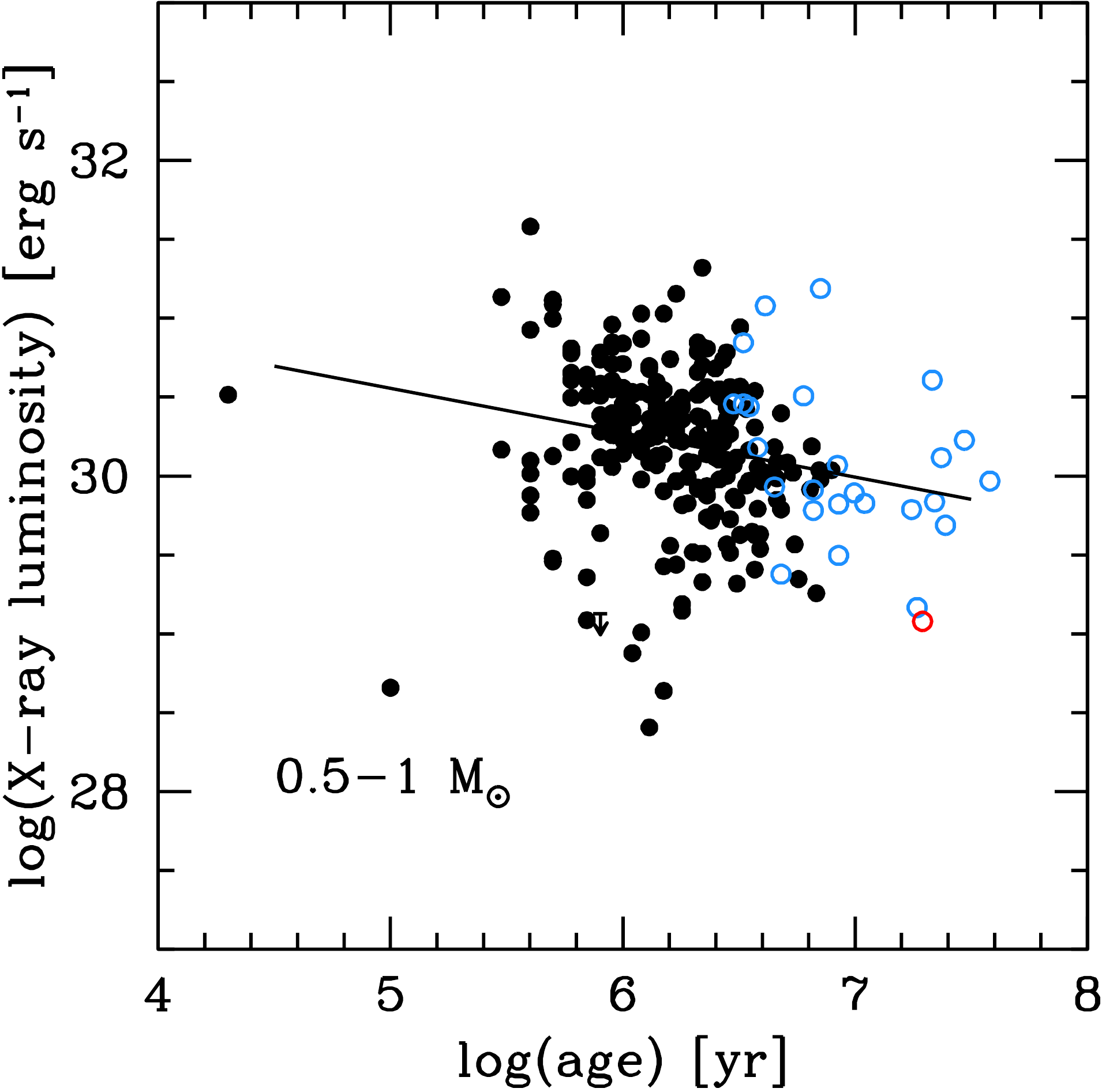} 
 \includegraphics[width=0.31\linewidth]{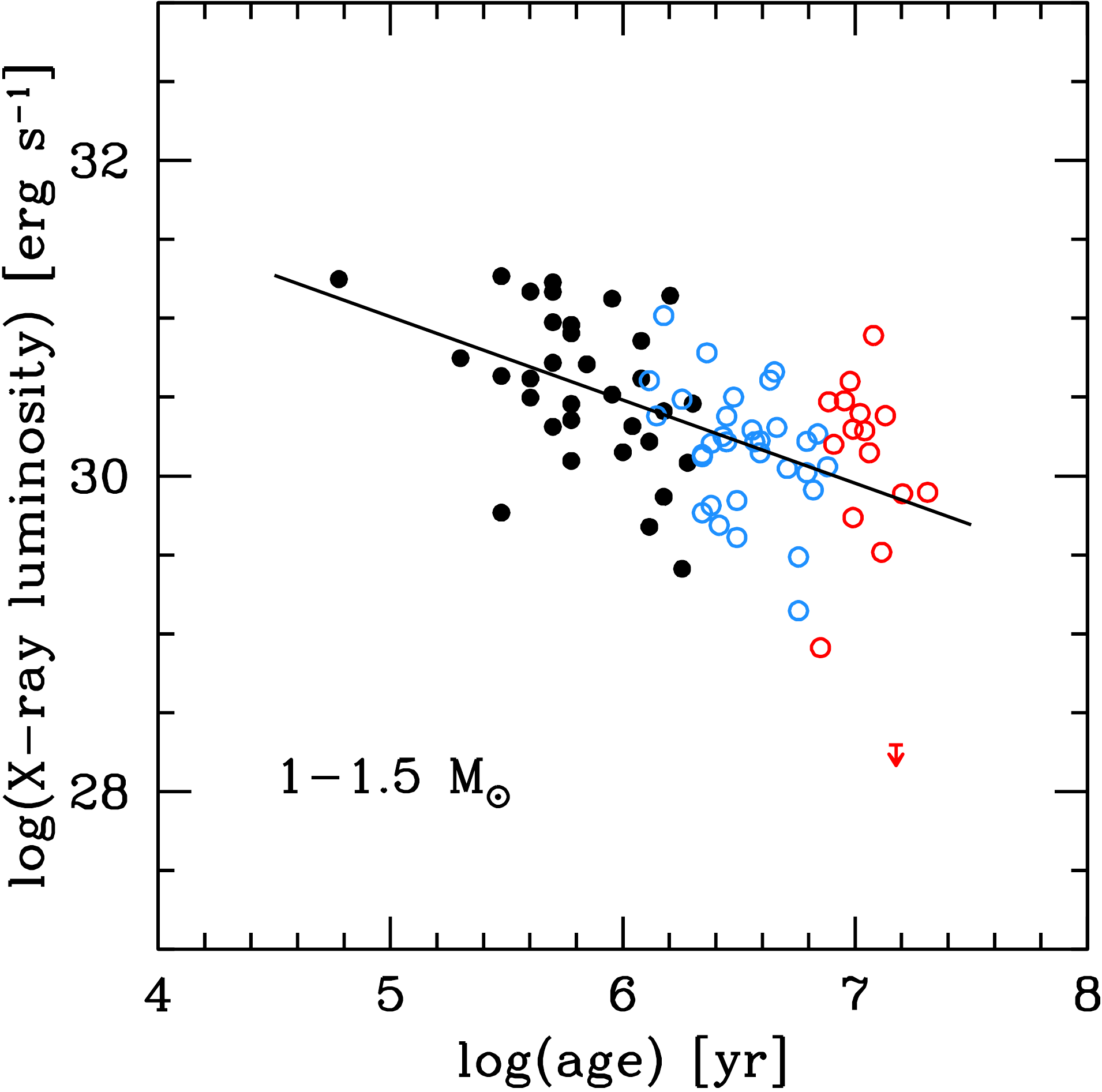}\\
 \includegraphics[width=0.31\linewidth]{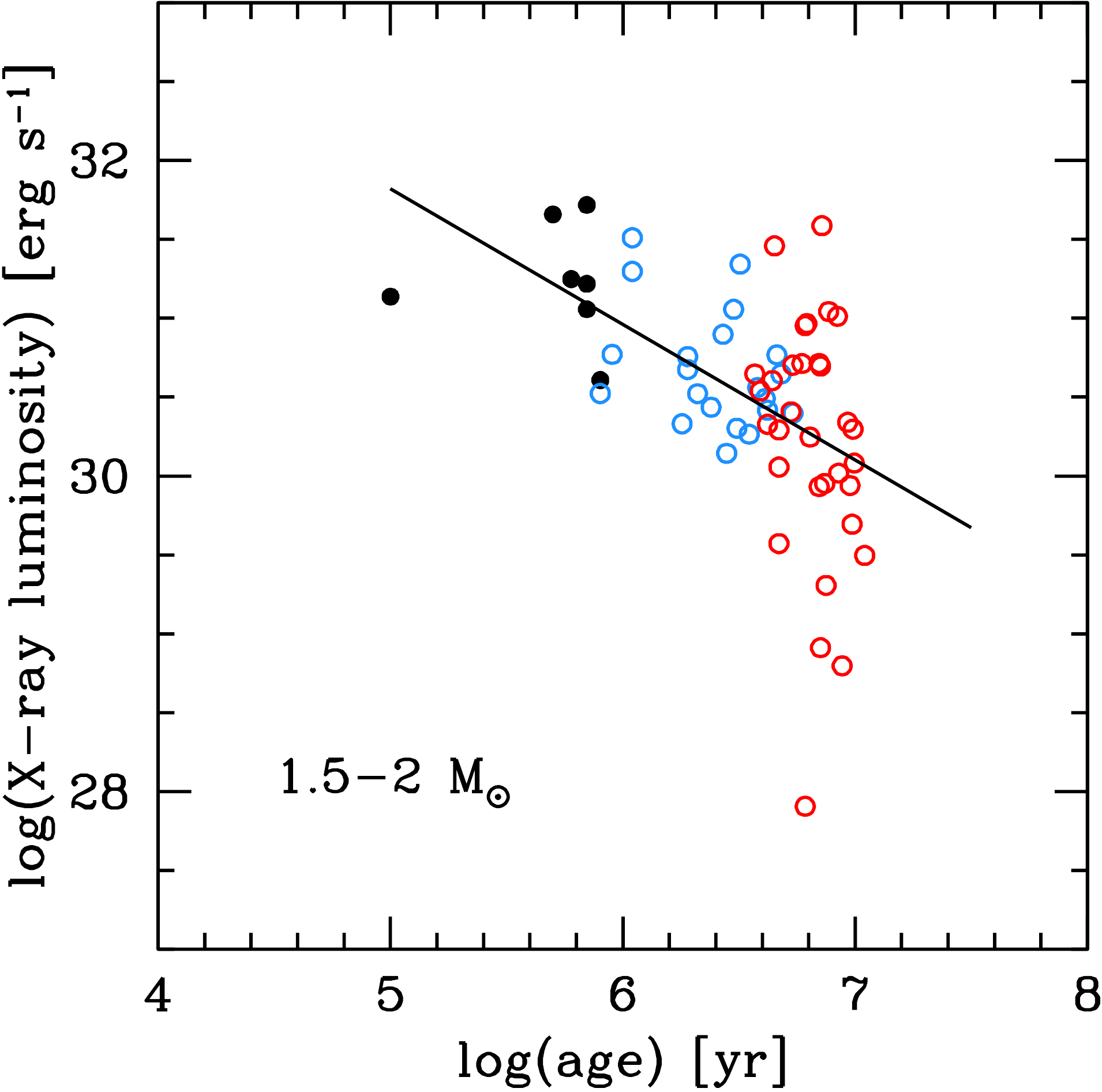}
 \includegraphics[width=0.31\linewidth]{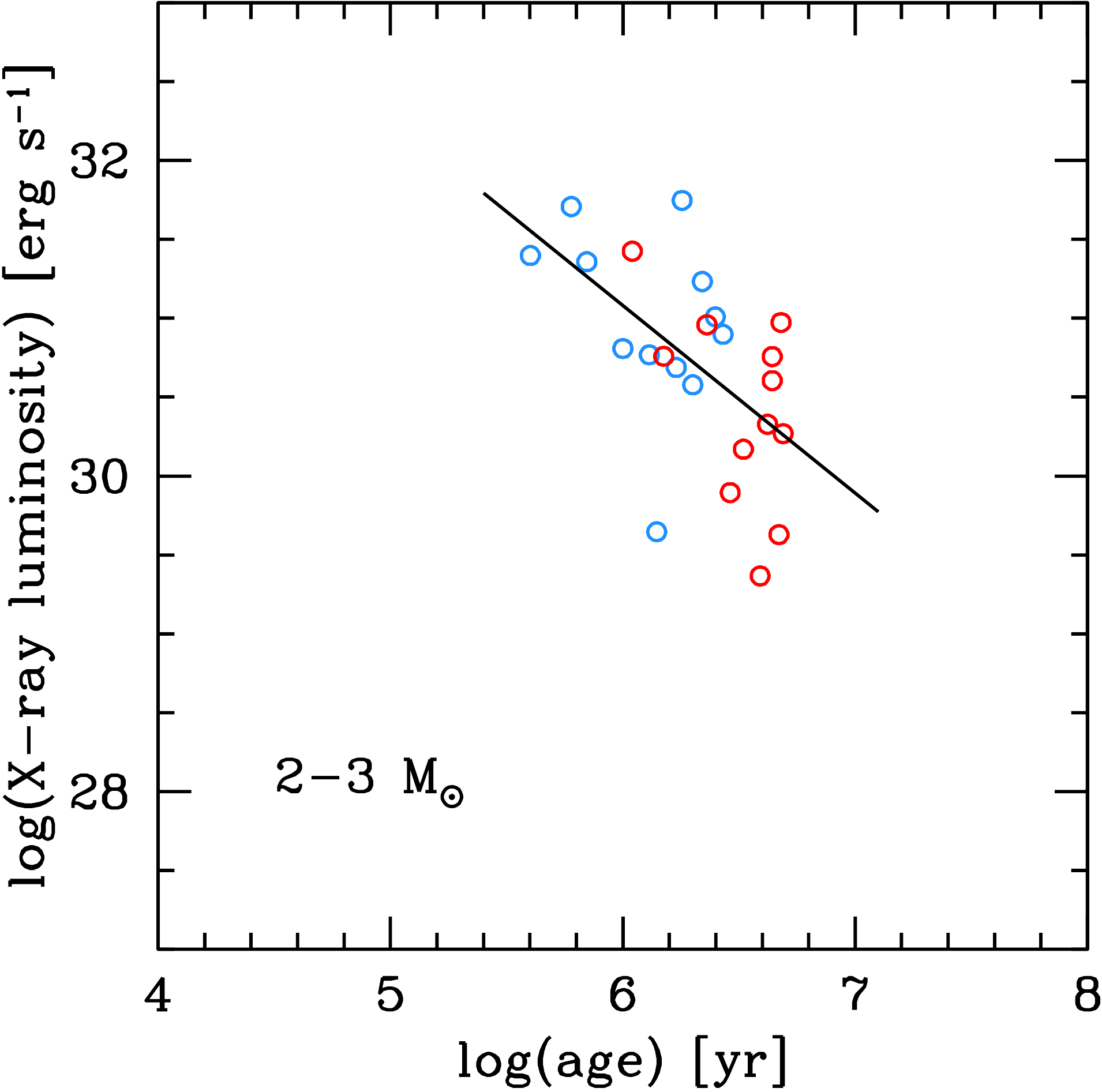}
 \caption{The decay of X-ray luminosity with age for PMS stars in specified mass bins (from \cite[Gregory et al. 2016]{gre16}). Black / blue / red points are fully convective / partially convective stars on Hayashi tracks / partially convective on Henyey tracks. $L_{\rm X}$ decays faster with age for stars in the higher mass bins, which have a greater proportion of partially convective stars.}
   \label{logLX_logage}
\end{center}
\end{figure}

Coronal X-ray emission decays with substantial radiative core development on the PMS (\cite[Gregory et al. 2016]{gre16}).  This is also apparent when considering the anti-correlation between $L_{\rm X}$ and stellar age, $t$, for PMS stars (see Figure \ref{logLX_logage}).  \cite[Preibisch \& Feigelson (2005)]{pre05b} reported that $L_{\rm X}\propto t^{-1/3}$ using a sample of mostly fully convective PMS stars from the ONC.  With a larger sample, and a substantial fraction of partially convective PMS stars, it is clear that the coronal X-ray emission decays faster with age for higher mass PMS stars (Figure \ref{logLX_logage}; \cite[Gregory et al. 2016]{gre16}).

The Henyey track PMS stars considered in Figures \ref{logLX_logLstar} \& \ref{logLX_logage} are currently early K-type and G-type stars (with a few F-types also).  These stars will evolve into MS A-type stars which lack outer convective zones (e.g. \cite[Siess et al. 2000]{sie00}). Some 85-90\% of MS A-types are undetected in X-rays, with almost all of those that are known / suspected close binaries (\cite[Schr{\"o}der \& Schmitt 2007]{sch07}).\footnote{A small number of single late A-types do show very weak levels of X-ray emission (e.g. \cite[Robrade \& Schmitt 2010]{rob10}; \cite[G{\"u}nther et al. 2012]{gun12}).}  The decay of the coronal X-ray emission when PMS stars become substantially radiative is consistent with the lack of X-ray emission from MS A-type stars.  Higher mass PMS stars that evolve onto Henyey tracks appear to lose their coronae as they develop large radiative cores, while lower mass PMS stars retain high levels of X-ray emission for millions of years of stellar evolution, as I discuss in the below.   


\vspace{-3.5mm}
\section{The long-term evolution coronal and chromospheric acitivty}\label{highenergy}
\begin{figure}[t]
\begin{center}
 \includegraphics[width=0.55\linewidth]{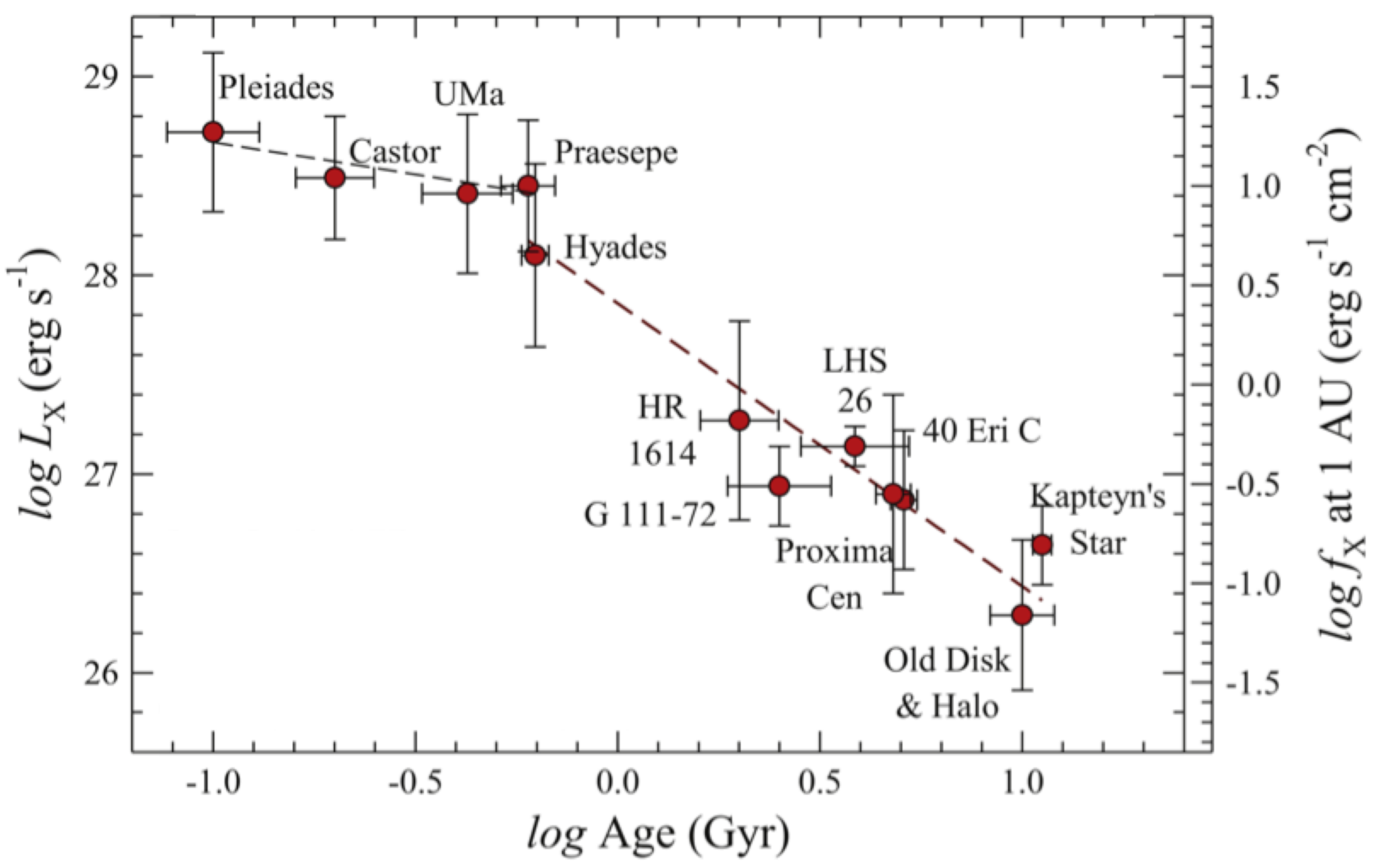} 
 \includegraphics[width=0.42\linewidth]{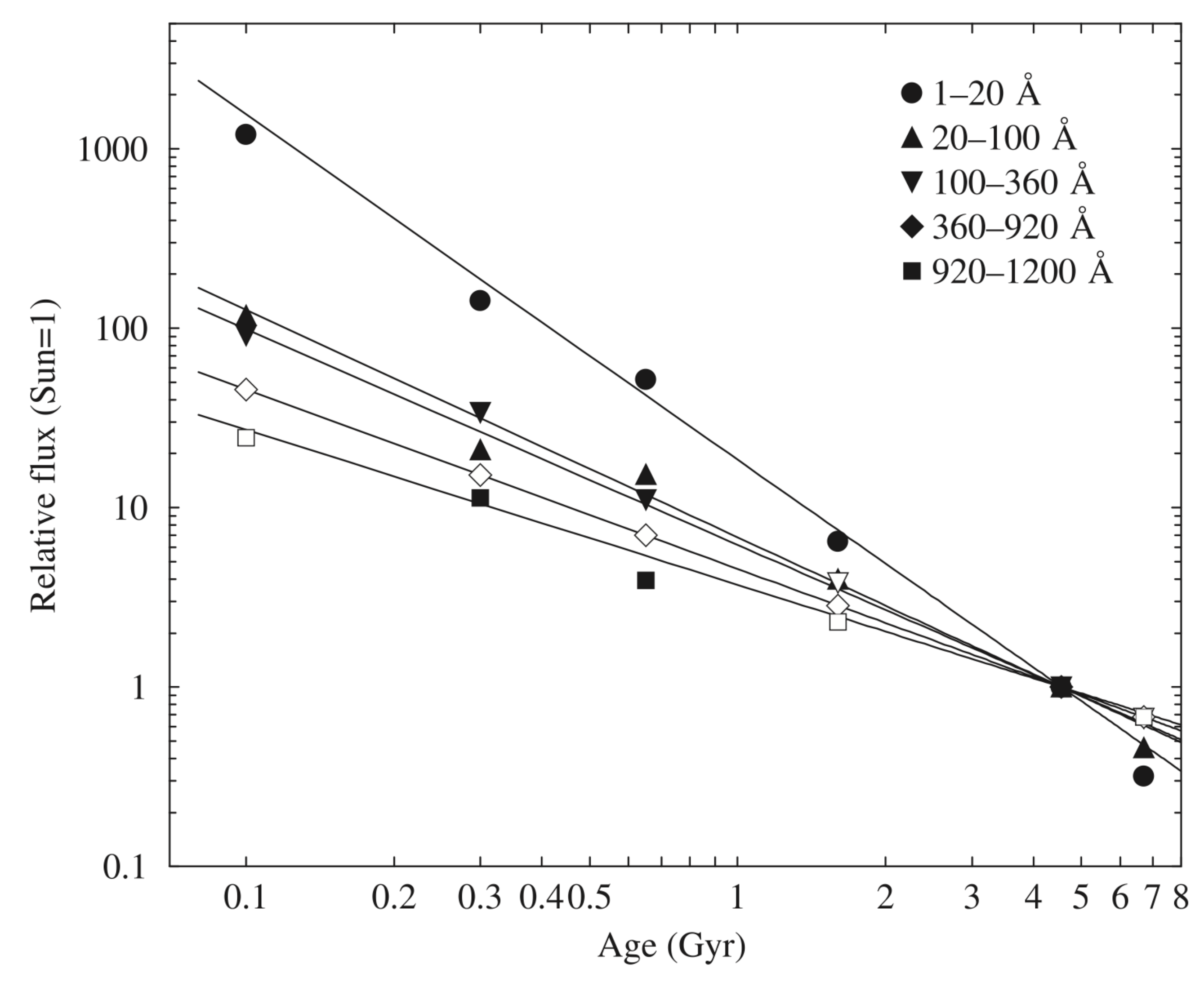} 
 \caption{(left) The decay of X-ray luminosity versus stellar age for M0-M5 stars. Figure from \cite[Guinan et al. (2016)]{gui16}. The right hand vertical axis shows the X-ray irradiance $f_{\rm X}$ at a distance of $1\,{\rm au}$. The dashed lines are least squares fits: for ${\rm age}>0.5\,{\rm Gyr}$, that is $\log({\rm age [Gyr]})>-0.3$, the fit is  $\log L_{\rm X}=27.857-1.424\log({\rm age [Gyr]})$.  X-ray activity decreases sharply with increasing age and as stars spin down (see \S\ref{rotation}). \copyright AAS. Reproduced with permission. (right) The decay of the (solar normalised) UV flux versus age. Figure from \cite[Ribas et al. (2005)]{rib05}.  Harder (shorter wavelength) emission reduces faster with age. \copyright AAS. Reproduced with permission.}
   \label{ribas}
\end{center}
\end{figure}

PMS stars are $10$-to-$10^4$ times more X-ray luminous than the Sun is today, and stars maintain high levels of X-ray emission for several $100\,{\rm Myr}$ of evolution before their activity begins to decay at a faster rate (see Figure \ref{ribas}, left panel). For ages beyond $\sim0.5\,{\rm Gyr}$ the X-ray luminosity decays with age as $L_{\rm X}\propto t^{-3/2}$ for both solar-type and lower mass stars (e.g. \cite[G{\"u}del et al. 1997]{gud97}; \cite[Guinan et al. 2016]{gui16}).  This corresponds to roughly the same age at which we observe convergence of the stellar rotation rates (see \S\ref{rotation} and Figure \ref{omega_age}).    

The UV emission also decays over Gyr timescales, although more slowly with age than the higher energy X-ray emission.  At FUV wavelengths, \cite[Guinan et al. (2016)]{gui16} report that the Ly$\alpha$ flux (a good proxy for FUV emission as a whole) decays with stellar age as roughly as $t^{-2/3}$ for early-to-mid spectral type M-dwarfs, more slowly than the $t^{-3/2}$ behaviour of the X-ray emission. As can be seen from Figure \ref{ribas} (right panel), the softer (longer wavelength) the emission the slower the decay with stellar age (e.g. \cite[Ribas et al. 2005]{rib05}; \cite[G{\"u}del 2007]{gud07}; \cite[Shkolnik \& Barman 2014]{shk14}).  Quantifying this further for solar analogues, \cite[G{\"u}del (2007)]{gud07} provides a list of enhancement factors of the high energy emission at various wavebands moving backwards in time from the present day solar values to those expected for the young Sun (see his table 6).  There is, however, considerable uncertainty in the activity level of the young Sun (\cite[Tu et al. 2015]{tu15}). Due to the large scatter in rotation rates at young cluster ages (see Figure \ref{rotation}) with the associated large scatter in activity levels, the rotation history of the Sun, and therefore of its activity, is unknown.      

\begin{figure}[ht]
\begin{center}
 \includegraphics[width=0.487\linewidth]{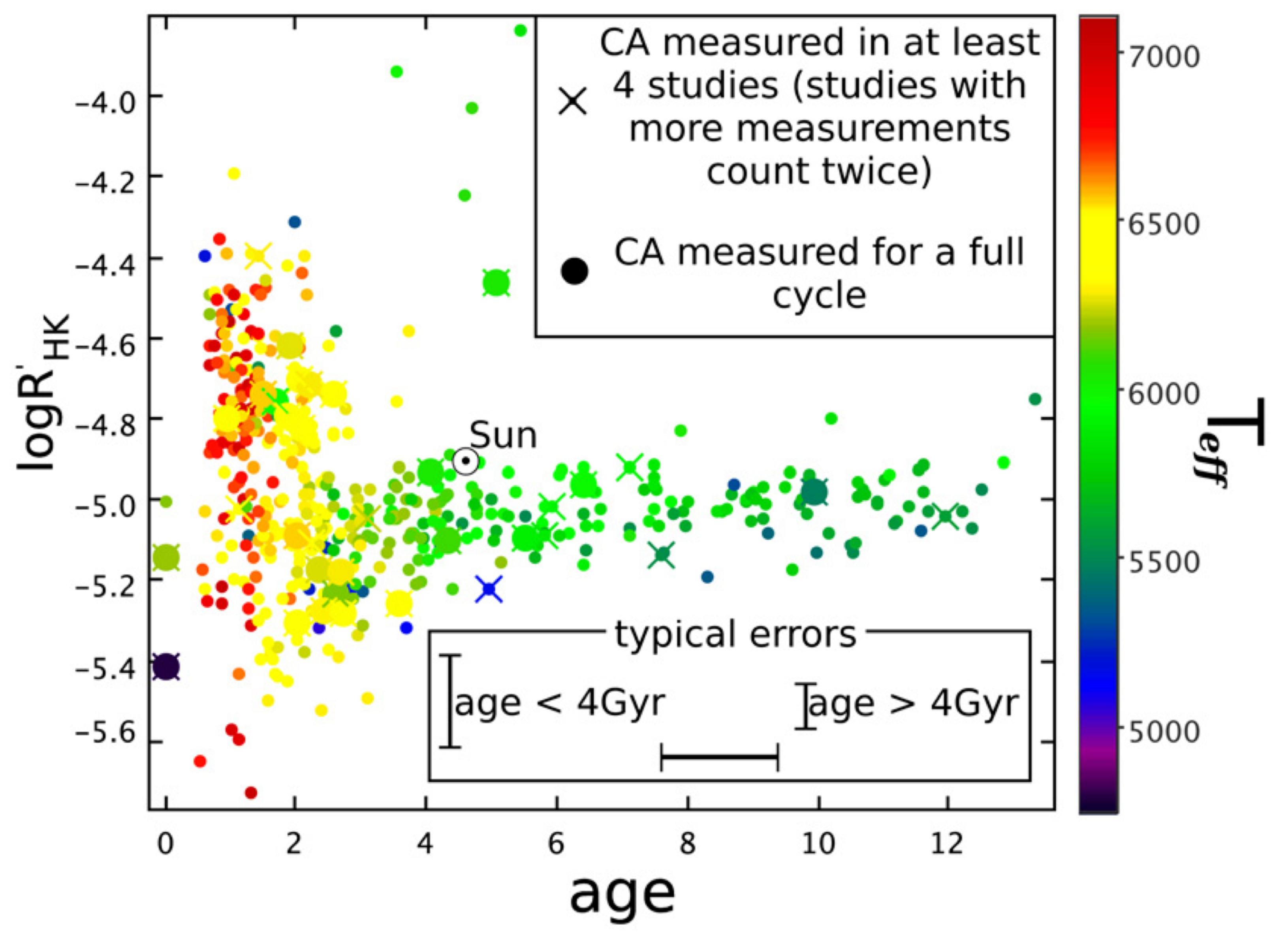} 
 \includegraphics[width=0.47\linewidth]{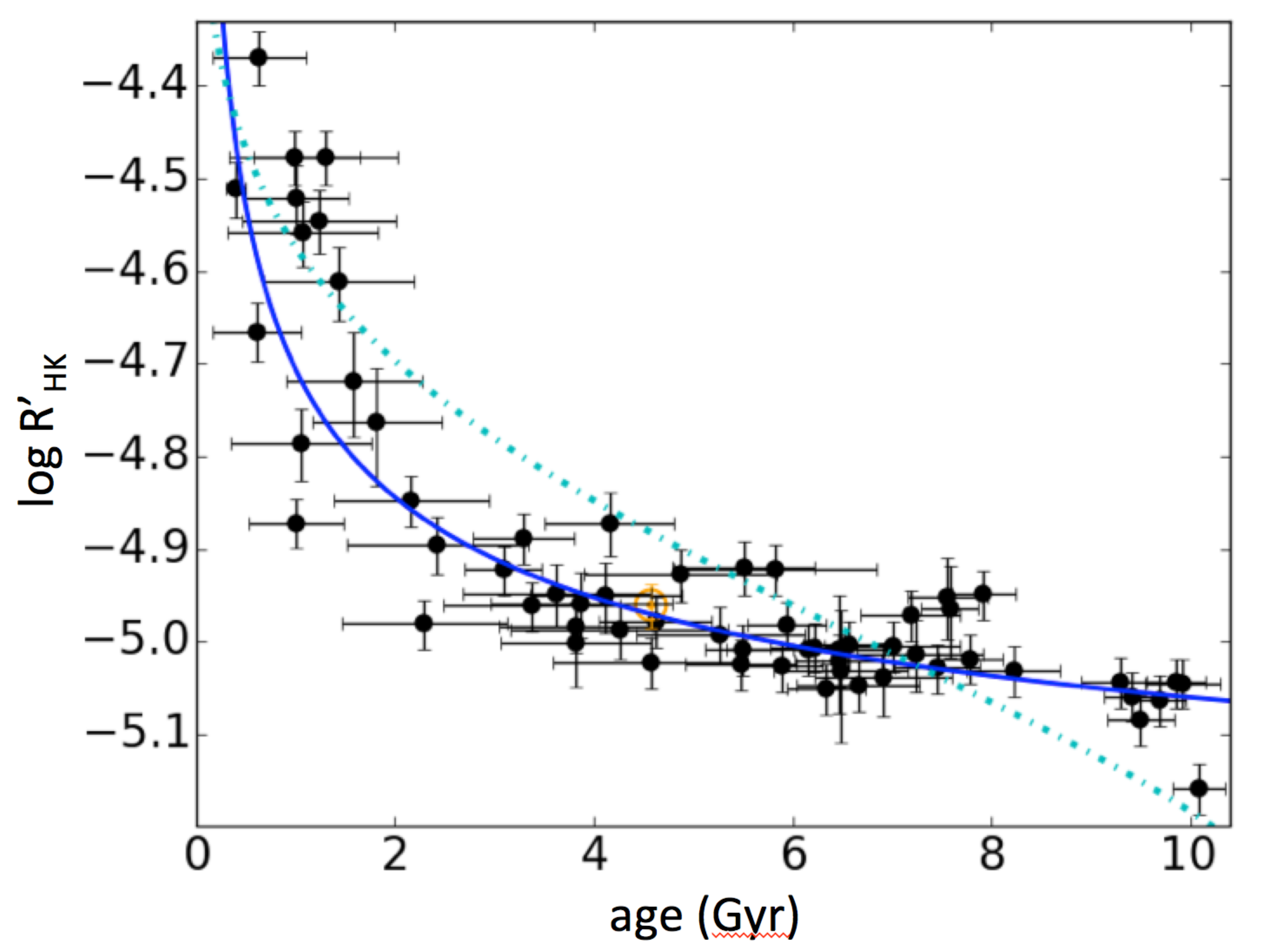} 
 \caption{The chromospheric CaII activity index $R'_{\rm HK}$ versus stellar age.  (left) Figure from \cite[Pace (2013)]{pac13} for stars with effective temperatures $T_{\rm eff}$ as indicated. Credit: Pace, A\&A, 551, L8, 2013, reproduced with permission \copyright ESO. (right) Figure modified from \cite[Freitas et al. (submitted)]{fre17}, see also Freitas et al., these proceedings, with the fit to the points shown as the solid blue line. The dashed line is the relationship derived by \cite[Mamajek \& Hillenbrand (2008)]{mam08}.}
   \label{freitas}
\end{center}
\end{figure}

Chromospheric activity as measured by $R'_{\rm HK}$ also decays over Gyr timescales. The chromospheric CaII activity index $R'_{\rm HK}$, see \cite[Noyes et al. (1984)]{noy84} for its introduction and definition, versus age is shown in Figure \ref{freitas}. $R'_{\rm HK}$ has been advocated as a proxy for stellar age (e.g. \cite[Mamajek \& Hillenbrand 2008]{mam08}), or at least for young ages (\cite[Pace 2013]{pac13}; Figure \ref{freitas} left panel) due to the flattening of the relation for older ages.

Recently, \cite[Freitas et al. (submitted)]{fre17} have re-examined the relationship between $R'_{\rm HK}$ and age, see Figure \ref{freitas} right panel and Freitas et al., these proceedings.  Studying CaII H \& K activity with HARPS spectra they conclude that \cite[Pace (2013)]{pac13} was right to argue that $R'_{\rm HK}$ is only a useful proxy for stellar age for younger clusters, with sensitivity to ages of $\lesssim3\,{\rm Gyr}$.  The \cite[Freitas et al. (submitted)]{fre17} data suggests that the continuing decrease in $R'_{\rm HK}$ with age reported by older studies (e.g. \cite[Mamajek \& Hillenbrand 2008]{mam08} - their relation is also plotted in Figure \ref{freitas} right panel for comparison) is incorrect, and that the chromospheric activity levels flatten off after $\sim3\,{\rm Gyr}$.  The flattening of the chromospheric activity at ages approaching and beyond solar age likely reflects the convergence of rotation rates during stellar MS evolution (see \S\ref{rotation} and Figure \ref{rotation}).  However, the flattening of the chromospheric activity may also be related to a recently reported flattening of rotation rates beyond $\sim2-3\,{\rm Gyr}$, which has been interpreted as being due to a break-down of the Skumanich spin-down law (e.g. \cite[dos Santos et al. 2016]{dos16}).       


\vspace{-3.5mm}
\section{The long-term evolution of stellar magnetism}\label{magnetism}
The indicators of stellar activity are driven by the interior dynamo and via the external magnetic field that it generates.  The past decade has seen major advances in the observational study of stellar magnetism, with magnetic field information now available for stars of all spectral types and evolutionary phases (e.g. \cite[Donati et al. 2006, 2008, 2011]{don06,don08,don11}; \cite[Morin et al. 2008]{mor08}; \cite[Petit et al. 2010]{pet10}; \cite[Fares et al. 2013]{far13}; \cite[Folsom et al. 2016]{fol16}). 

On the PMS, stars of mass $\gtrsim0.5\,{\rm M}_\odot$ appear to be born with simple, axisymmetric, often dominantly octupolar large-scale magnetic fields, that transition to complex and non-axisymmetric with significant radiative core development (\cite[Gregory et al. 2012]{gre12}; \cite[Folsom et al. 2016]{fol16}). PMS stars of lower mass have been predicted to host a variety of magnetic field topologies based on the similarities of PMS magnetism and the field topologies found for MS M-dwarfs with similar internal structures (\cite[Gregory et al. 2012]{gre12}). The lowest mass MS M-dwarfs host a mixture of simple and complex magnetic fields and exist in a bistable dynamo regime (\cite[Morin et al. 2010]{mor10}).  The lowest mass PMS stars may follow similar behaviour, but nIR spectropolarimetry (with, for example, the SPIRou instrument; \cite[Moutou et al. 2015]{mou15}) is required to properly survey their magnetic fields. 

For MS stars, snapshot surveys have confirmed that stars with deeper convective zones (K-types) have stronger average longitudinal magnetic field components, $\langle|B_\ell |\rangle$, than those with smaller convective envelopes (F and G-types; \cite[Marsden et al. 2014]{mar14}). This, however, may be due to the selection of active K-type stars for the survey, or a flux cancellation effect if the F and G-types have more numerous and smaller magnetic spots than K-types. $\langle|B_\ell |\rangle$ increases with rotation rate, decreases with age, and is well correlated with the chromospheric activity (\cite[Marsden et al. 2014]{mar14}).  

\begin{figure}[t]
\begin{center}
 \includegraphics[width=0.52\linewidth]{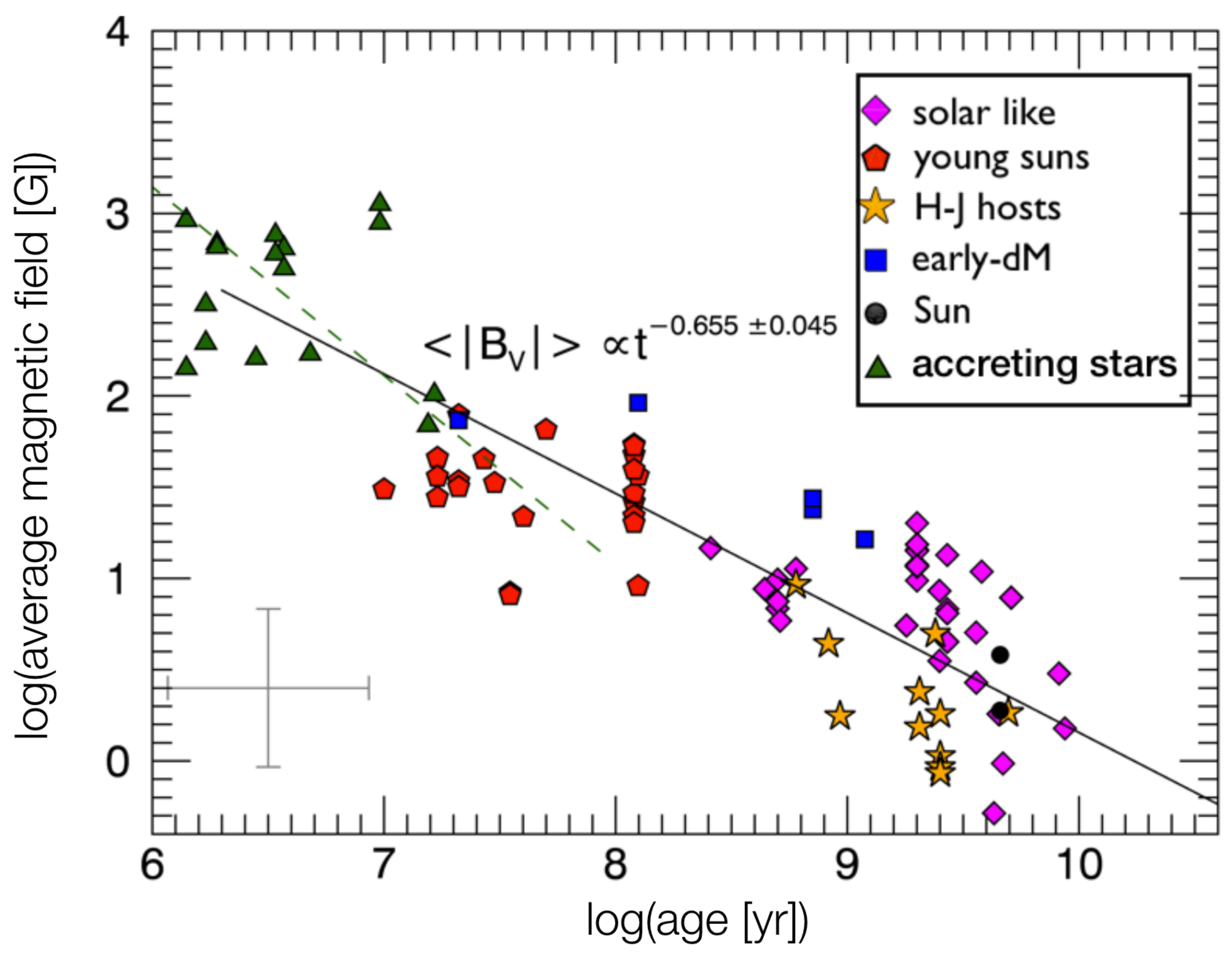} 
 \caption{The decay of the mean magnetic field strength measured from magnetic maps derived from Zeeman-Doppler imaging for stars of various age / evolutionary stage. Combined with the Skumanich spin-down law, $\Omega_\ast\propto t^{-1/2}$, this is roughly consistent with the stellar dynamos being linear ($\langle B\rangle\propto\Omega_\ast$). Figure modified from \cite[Vidotto et al. (2014)]{vid14}.}
   \label{vidotto}
\end{center}
\end{figure}

Zeeman-Doppler imaging studies, which allow magnetic maps to be constructed from a time series of circularly polarised spectra, have revealed how the magnetic topologies of stars are linked to the stellar parameters - see the reviews by \cite[Donati (2013)]{don13} and \cite[Morin (2012)]{mor12}.  Using almost all of the stellar magnetic maps of low-mass stars available at the time, \cite[Vidotto et al. (2014)]{vid14} examined trends in the large-scale stellar magnetism with age.  Multiple correlations between age, activity, and rotation parameters were reported including that shown in Figure \ref{vidotto}.  Over Gyr timescales the average unsigned magnetic field strength of FGKM-type stars, as measured from magnetic maps, decays with age as $\langle|B|\rangle\propto t^{-0.66\pm0.05}$.  This, combined with the Skumanich spin-down law, confirms that stellar dynamos are close to being linear.     

\vspace{-3.5mm}
\section{Summary: key points}\label{summary}
\begin{itemize}
\item There is a large distribution of rotation rates of stars in young clusters. Rotation rates converge for ages $\gtrsim0.5\,{\rm Gyr}$ (e.g. \cite[Gallet \& Bouvier 2015]{gal15}).  Following the PMS evolution, main sequence stars spin-down via angular momentum loss in magnetised winds with the faster rotators spinning down at a faster rate than the slower rotators. 
\vspace{1.5mm}
\item Over Gyr timescales, low-mass stars spin down as $\Omega_\ast\propto t^{-1/2}$, the Skumanich spin-down law (\cite[Skumanich 1972]{sku72}). However, there are recent suggestions this relation breaks down, with the rotation rate decay flattening for ages $\gtrsim$3$\,{\rm Gyr}$ (\cite[dos Santos et al. 2016]{dos16}).
\vspace{1.5mm}
\item Many stellar activity diagnostics, e.g. X-ray emission, flare activity, H$\alpha$ activity, chromospheric CaII H \& K emission, and others, follow a rotation-activity relation, whereby activity increases with increasing rotation rate, equivalently with decreasing Rossby number, Ro. Activity saturates for the fastest rotators.  For example, X-ray activity saturates at $L_{\rm X}/L_\ast\approx 10^{-3}$ for ${\rm Ro}\lesssim0.1$ (e.g. \cite[Wright et al. 2011]{wri11}).  
\vspace{1.5mm}   
\item X-ray activity saturates at ${\rm Ro}=P_{\rm rot}/\tau_{\rm c}\lesssim0.1$ regardless of spectral type (e.g. \cite[Vilhu 1984]{vil84}).  Later spectral types have longer turnover times, $\tau_{\rm c}$, than for earlier spectral types. Therefore, fully convective M-dwarfs ($\sim$M3.5 and later) display saturated emission.  However, \cite[Wright et al. (2016)]{wri16}, have recently shown that M-dwarfs that have sufficiently spun down can fall in the unsaturated regime. This means that a tachocline is not a requirement to generate magnetic fields that link the activity level with rotation rate.
\vspace{1.5mm}
\item Stars in young PMS clusters (e.g. the ONC) do not follow the MS rotation-activity relation. All young PMS stars show saturated levels of X-ray activity but with orders of magnitude more scatter in $L_{\rm X}/L_\ast$ (\cite[Preibisch et al. 2005]{pre05a}). 
\vspace{1.5mm}
\item The youngest region where stars (at least those of mass $1<M_\ast/{\rm M}_\odot<1.4$) have begun to follow MS-like rotation-activity behaviour is the $\sim13\,{\rm Myr}$ old cluster h~Per (\cite[Argiroffi et al. 2016]{arg16}).  However, the gradient of the unsaturated regime in h~Per is less than found for stars in MS clusters.  The emergence of unsaturated behaviour may be driven by stars developing substantial radiative cores.  
\vspace{1.5mm}
\item PMS stars (at least those of mass $\gtrsim0.5\,{\rm M}_\odot$) are born with simple axisymmetric magnetic fields that evolve to dominantly non-axisymmetric and highly multipolar with substantial radiative core development (\cite[Gregory et al. 2012]{gre122016}). Lower mass PMS stars may exist in a bistable dynamo regime and host a variety of large-scale magnetic field topologies (\cite[Gregory et al. 2012]{gre12}). Additional observations are required for confirmation.   
\vspace{1.5mm}
\item PMS stars that have developed substantial radiative cores, those which have evolved onto Henyey tracks in the H-R diagram, have lower fractional X-ray luminosities ($L_{\rm X}/L_\ast$) than those on Hayashi tracks (\cite[Rebull et al. 2006]{reb06}; \cite[Gregory et al. 2016]{gre16}).  Many such PMS stars are the progenitors of MS A-type stars, and the loss of their coronal X-ray emission is consistent with the lack of X-ray detections of MS A-types (\cite[Gregory et al. 2016]{gre16}). 
\vspace{1.5mm}
\item Over Gyr timescales, surface average magnetic field strengths reduce with age as $\langle|B|\rangle\propto t^{-2/3}$ (\cite[Vidotto et al. 2014]{vid14}). Snapshot surveys of MS stars have shown that the average longitudinal magnetic field component, $\langle|B_\ell |\rangle$, is well correlated with chromospheric activity (\cite[Marsden et al. 2014]{mar14}). 
\vspace{1.5mm}
\item Over Gyr timescales, harder emission decays faster with age than softer emission (\cite[Ribas et al. 2005]{rib05}; \cite[G{\"u}del 2007]{gue07}).  For example, for ages $\gtrsim0.5\,{\rm Gyr}$ the X-ray emission decays faster with age, $\propto t^{-3/2}$, than the FUV emission, $\propto t^{-2/3}$ (\cite[Guinan et al. 2016]{gui16}).
\vspace{1.5mm}
\item The chromospheric CaII activity index $R'_{\rm HK}$ decays rapidly with, and is a useful proxy for, stellar age for $t\lesssim3\,{\rm Gyr}$ (\cite[Pace 2013]{pac13}; \cite[Freitas et al. submitted]{fre17}).  The relationship between $R'_{HK}$ and $t$ is almost flat for older ages.  
\end{itemize}


\vspace{-3.5mm}
\section*{Acknowledgements}
I acknowledge support from the Science \& Technology Facilities Council (STFC) via an Ernest Rutherford Fellowship [ST/J003255/1] and the meeting organisers for the invitation to speak. I thank Nick Wright \& Stephen Marsden for comments on their work and Fabr{\'i}cio Freitas for kindly providing and modifying Figure \ref{freitas} (right panel). 




\vspace{-3.5mm}
\begin{thebibliography}{}

\bibitem[Alexander \& Preibisch (2012)]{ale12}
{Alexander, F., \& Preibisch, T.} 2012, 
\textit{A\&A}, 539, A64 

\bibitem[Amelin et al. (2010)]{ame10}
{Amelin, Y., Kaltenbach, A., Iizuka, T., et al.} 2010, 
\textit{Earth and Planetary Science Letters}, 300, 343 

\bibitem[Argiroffi et al. (2016)]{arg16}
{Argiroffi, C., Caramazza, M., Micela, G., et al.} 2016, 
\textit{A\&A}, 589, A113

\bibitem[Audard et al. (2000)]{aud00}
{Audard, M., G{\"u}del, M., Drake, J.J., \& Kashyap, V.L.} 2000,
\textit{ApJ}, 541, 396 

\bibitem[Bahcall et al. (1995)]{bah95} 
{Bahcall, J.N., Pinsonneault, M.H., \& Wasserburg, G.J.} 1995, 
\textit{Reviews of Modern Physics}, 67, 781 

\bibitem[Broos et al. (2013)]{bro13}
{Broos, P.S., Getman, K.V., Povich, M.S., et al.} 2013, 
\textit{ApJS}, 209, 32 

\bibitem[Chabrier \& Baraffe (1997)]{cha97}
{Chabrier, G., \& Baraffe, I.} 1997, 
\textit{A\&A}, 327, 1039 

\bibitem[Davenport (2016)]{dav16}
{Davenport, J.R.A.} 2016,
\textit{ApJ}, 829, 23 

\bibitem[Davies et al. (2014)]{dav14}
{Davies, C.L., Gregory, S.G., \& Greaves, J.S} 2014, 
\textit{MNRAS}, 444, 1157 

\bibitem[Donati (2013)]{don13}{Donati, J.-F.} 2013, 
EAS Publications Series, 62, 289 

\bibitem[Donati et al. (2011)]{don11}{Donati, J.-F., Gregory, S.G., Alencar, S.H.P., et al.} 2011, 
\textit{MNRAS}, 417, 472 

\bibitem[Donati et al. (2008)]{don08}{Donati, J.-F., Morin, J., Petit, P., et al.} 2008,
\textit{MNRAS}, 390, 545 

\bibitem[Donati et al. (2006)]{don06}
{Donati, J.-F., Howarth, I.D., Jardine, M.M., et al.} 2006, 
\textit{MNRAS}, 370, 629 

\bibitem[Douglas et al. (2014)]{dou14}
{Douglas, S.T., Ag{\"u}eros, M.A., Covey, K.R., et al.} 2014,
\textit{ApJ}, 795, 161 

\bibitem[dos Santos et al. (2016)]{dos16}
{dos Santos, L.A., Mel{\'e}ndez, J., do Nascimento, J.-D., et al.} 2016,
\textit{A\&A}, 592, A156 

\bibitem[Dunham \& Vorobyov (2012)]{dun12}
{Dunham, M.M., \& Vorobyov, E.I.} 2012, 
\textit{ApJ}, 747, 52  

\bibitem[Fares et al. (2013)]{far13}
{Fares, R., Moutou, C., Donati, J.-F., et al.} 2013,
\textit{MNRAS}, 435, 1451 

\bibitem[Folsom et al. (2016)]{fol16}
{Folsom, C.P., Petit, P., Bouvier, J., et al.} 2016,
\textit{MNRAS}, 457, 580 

\bibitem[Flaccomio et al. (2005)]{fla05}
{Flaccomio, E., Micela, G., Sciortino, S., et al.} 2005,
\textit{ApJS}, 160, 450 

\bibitem[Freitas et al. (submitted)]{fre17}
{Freitas, F.C., Mel{\'e}ndez, J., Bedell, M., et al.} 2017,
\textit{A\&A}, submitted

\bibitem[Gallet \& Bouvier (2015)]{gal15}
{Gallet, F., \& Bouvier, J.} 2015, 
\textit{A\&A}, 577, A98

\bibitem[Gerber et al. (2010)]{ger10}
{Gerber, R.J., Wilks T., \& Erdie-Lalena, C.} 2010,
\textit{Pediatrics in Review}, 31, 267 

\bibitem[Gregory et al. (2016)]{gre16}
{Gregory, S.G., Adams, F.C., \& Davies, C.L.} 2016, 
\textit{MNRAS}, 457, 3836 

\bibitem[Gregory et al. (2012)]{gre12}
{Gregory, S.G., Donati, J.-F., Morin, J., et al.} 2012, 
\textit{ApJ}, 755, 97 

\bibitem[G{\"u}del (2007)]{gud07}
{G{\"u}del, M.} 2007, 
\textit{Living Reviews in Solar Physics}, 4, 3 

\bibitem[G{\"u}del et al. (1997)]{gud97}
{G{\"u}del, M., Guinan, E.F., \& Skinner, S.L.} 1997,
\textit{ApJ}, 483, 947 

\bibitem[Guinan et al. (2016)]{gui16}
{Guinan, E.F., Engle, S.G., \& Durbin, A.} 2016,
\textit{ApJ}, 821, 81 

\bibitem[G{\"u}nther et al. (2012)]{gun12}
{G{\"u}nther, H.M., Wolk, S.J., Drake, J.J., et al.} 2012,
\textit{ApJ}, 750, 78 

\bibitem[Haisch et al. (2001)]{hai01} 
{Haisch, K.E., Jr., Lada, E.A., \& Lada, C.J.} 2001, 
\textit{ApJL}, 553, L153 

\bibitem[James et al. (2000)]{jam00}
{James, D.J., Jardine, M.M., Jeffries, R.D., et al.} 2000, 
\textit{MNRAS}, 318, 1217 

\bibitem[Jardine \& Unruh (1999)]{jar99}
{Jardine, M., \& Unruh, Y.C.} 1999, 
\textit{A\&A}, 346, 883 

\bibitem[Jardine (2004)]{jar04}
{Jardine, M.} 2004, 
\textit{A\&A}, 414, L5 

\bibitem[Jeffries et al. (2011)]{jef11} 
{Jeffries, R.D., Jackson, R.J., Briggs, K.R., Evans, P.A., \& Pye, J.P.}2011, 
\textit{MNRAS}, 411, 2099 

\bibitem[Kiraga \& Stepien (2007)]{kir07}
{Kiraga, M., \& Stepien, K.} 2007, 
\textit{Acta Astronomica}, 57, 149 

\bibitem[Landin et al. (2010)]{lan10}
{Landin, N.R., Mendes, L.T.S., \& Vaz, L.P.R.} 2010, 
\textit{A\&A}, 510, A46 

\bibitem[Maeder (2009)]{mae09}
{Maeder, A.} 2009, 
\textit{Physics, Formation and Evolution of Rotating Stars}, Astronomy and Astrophysics Library, Springer Berlin Heidelberg  

\bibitem[Mamajek \& Hillenbrand (2008)]{mam08}
{Mamajek, E.E., \& Hillenbrand, L.A.} 2008, 
\textit{ApJ}, 687, 1264 

\bibitem[Marsden et al. (2014)]{mar14}
{Marsden, S.C., Petit, P., Jeffers, S.V., et al.} 2014, 
\textit{MNRAS}, 444, 3517 

\bibitem[Morin (2012)]{mor12}
{Morin, J.} 2012, 
EAS Publications Series, 57, 165 

\bibitem[Morin et al. (2010)]{mor10}
{Morin, J., Donati, J.-F., Petit, P., et al.} 2010, 
\textit{MNRAS}, 407, 2269 

\bibitem[Morin et al. (2008)]{mor08}
{Morin, J., Donati, J.-F., Petit, P., et al.} 2008, 
\textit{MNRAS}, 390, 567 

\bibitem[Moutou et al. (2015)]{mou15}{Moutou, C., Boisse, I., H{\'e}brard, G., et al.} 2015, 
SF2A-2015: Proceedings of the Annual meeting of the French Society of Astronomy and Astrophysics, 205

\bibitem[Newton et al. (2016)]{new16}
{Newton, E.R., Irwin, J., Charbonneau, D., et al.} 2016, 
\textit{ApJ}, in press [astro-ph/1611.03509] 

\bibitem[Noyes et al. (1984)]{noy84}
{Noyes, R.W., Hartmann, L.W., Baliunas, S.L., Duncan, D.K., \& Vaughan, A.H.} 1984, 
\textit{ApJ}, 279, 763  

\bibitem[Offner \& McKee (2011)]{off11} 
{Offner, S.S.R., \& McKee, C.F.} 2011, 
\textit{ApJ}, 736, 53 

\bibitem[Pace (2013)]{pac13}
{Pace, G.} 2013, 
\textit{A\&A}, 551, L8

\bibitem[Pallavicini et al. (1981)]{pal81}
{Pallavicini, R., Golub, L., Rosner, R., et al.} 1981, 
\textit{ApJ}, 248, 279

\bibitem[Petit et al. (2010)]{pet10}
{Petit, P., Ligni{\`e}res, F., Wade, G.A., et al.} 2010,
\textit{A\&A}, 523, A41 

\bibitem[Pizzolato et al. (2003)]{piz03}
{Pizzolato, N., Maggio, A., Micela, G., Sciortino, S., \& Ventura, P.} 2003, 
\textit{A\&A}, 397, 147 

\bibitem[Pols et al. (1998)]{pol98}
{Pols, O.R., Schr{\"o}der, K.-P., Hurley, J.R., Tout, C.A., \& Eggleton, P.P.} 1998,
\textit{MNRAS}, 298, 525 

\bibitem[Preibisch et al. (2005)]{pre05a}
{Preibisch, T., Kim, Y.-C., Favata, F., et al.} 2005, 
\textit{ApJS}, 160, 401

\bibitem[Preibisch \& Feigelson (2005)]{pre05b}
{Preibisch, T., \& Feigelson, E.D.} 2005, 
\textit{ApJS}, 160, 390 

\bibitem[Rebull et al. (2006)]{reb06}
{Rebull, L.M., Stauffer, J.R., Ramirez, S.V., et al.} 2006, 
\textit{AJ}, 131, 2934

\bibitem[Reiners et al. (2014)]{rei14}
{Reiners, A., Sch{\"u}ssler, M., \& Passegger, V.M.} 2014,
\textit{ApJ}, 794, 144 

\bibitem[Reiners et al. (2009)]{rei09}
{Reiners, A., Basri, G., \& Browning, M.} 2009,
\textit{ApJ}, 692, 538 

\bibitem[Ribas et al. (2005)]{rib05}
{Ribas, I., Guinan, E.F., G{\"u}del, M., \& Audard, M.} 2005, 
\textit{ApJ}, 622, 680 

\bibitem[Robrade \& Schmitt(2010)]{rob10}
{Robrade, J., \& Schmitt, J.H.M.M.} 2010,
\textit{A\&A}, 516, A38

\bibitem[Schr{\"o}der \& Schmitt (2007)]{sch07}
{Schr{\"o}der, C., \& Schmitt, J.H.M.M.} 2007, 
\textit{A\&A}, 475, 677 

\bibitem[Shkolnik \& Barman (2014)]{shk14}
{Shkolnik, E.L., \& Barman, T.S.} 2014,
\textit{AJ}, 148, 64 

\bibitem[Siess et al. (2000)]{sie00}
{Siess, L., Dufour, E., \& Forestini, M.} 2000, 
\textit{A\&A}, 358, 593

\bibitem[Skumanich (1972)]{sku72}
{Skumanich, A.} 1972, 
\textit{ApJ}, 171, 565

\bibitem[Soderblom et al. (1993)]{sod93}
{Soderblom, D.R., Stauffer, J.R., Hudon, J.D., \& Jones, B.F.} 1993, 
\textit{ApJS}, 85, 315 

\bibitem[Stelzer et al. (2012)]{ste12}
{Stelzer, B., Preibisch, T., Alexander, F., et al.} 2012,
\textit{A\&A}, 537, A135 

\bibitem[Tu et al. (2015)]{tu15}
{Tu, L., Johnstone, C.P., G{\"u}del, M., \& Lammer, H.} 2015,
\textit{A\&A}, 577, L3 

\bibitem[Vidotto et al. (2014)]{vid14}
{Vidotto, A.A., Gregory, S.G., Jardine, M., et al.} 2014, 
\textit{MNRAS}, 441, 2361  

\bibitem[Vilhu (1984)]{vil84}
{Vilhu, O.} 1984,
\textit{A\&A}, 133, 117 

\bibitem[Wright \& Drake (2016)]{wri16}
{Wright, N.J., \& Drake, J.J.} 2016, 
\textit{Nature}, 535, 526 

\bibitem[Wright et al. (2011)]{wri11}
{Wright, N.J., Drake, J.J., Mamajek, E.E., \& Henry, G.W.} 2011, 
\textit{ApJ}, 743, 48 

\end{thebibliography}
\end{document}